\documentclass[
reprint, superscriptaddress, amsmath,amssymb, aps, ]{revtex4-2}

\usepackage{graphicx}
\usepackage{dcolumn}
\usepackage{bm}

\usepackage{placeins}

\newcommand{\rev}[1]{#1}

\begin{document}


\title{Impedance of multilayer cylindrical structures with a material-filled beam region}

\author{Erik Kvikne}
\thanks{Contact author: erik.kvikne@cern.ch}
\affiliation{European Organization for Nuclear Research (CERN), Geneva, Switzerland}
\affiliation{Department of Physics, University of Oslo, Oslo, Norway}

\author{Elias M\'etral}
\affiliation{European Organization for Nuclear Research (CERN), Geneva, Switzerland}

\author{Erik Adli}
\affiliation{Department of Physics, University of Oslo, Oslo, Norway}

\date{\today}

\begin{abstract}
Beam-coupling impedances in material media are relevant for ionization-cooling channels of a future muon collider, where the beam propagates in matter rather than vacuum. We extend the cylindrical field-matching formalism for multilayer structures to the case of a material-filled beam region surrounded by external layers with arbitrary electromagnetic properties. Relative to the vacuum formulation, the usual factor $1/\gamma^2$ is replaced by the material-dependent factor $F=1/\varepsilon_1-\mu_1\beta^2$, and the radial propagation constant is modified accordingly. Analytical expressions are obtained for the monopolar longitudinal and dipolar transverse impedances, with the surrounding structure encoded through reflection coefficients determined by field matching. The formalism reduces to the known vacuum and perfectly conducting limits in the appropriate cases. Representative calculations are presented for absorber-relevant configurations, illustrating the dependence on the material properties of both the beam region and the surrounding layers. In particular, a finite beam-region conductivity can generate a real impedance component and reverse the sign of the imaginary space-charge impedance, while sufficiently high permittivity can give rise to resonant structures associated with oscillatory radial fields. The effect of finite-conductivity surroundings is also examined, showing the approach towards the perfectly conducting limit as the external conductivity is increased. The applicability of the infinite-length approximation is also discussed by comparison with finite-length mode-matching results.
\end{abstract}

\maketitle

\section{Introduction}

Beam-coupling impedances are a central ingredient in the description of collective effects in high-intensity accelerators. In particular, the longitudinal and transverse resistive-wall impedances have been studied for decades because of their important role in beam stability, tune shifts, and emittance degradation. In their classical form, resistive-wall models describe a beam propagating in vacuum inside a conducting beam pipe, and a large body of work has established analytical expressions for a variety of geometries and material regimes. In the last couple of decades, renewed interest in this subject emerged in connection with accelerator components having small apertures and thick surrounding walls in machines with large circumferences, such as the CERN LHC collimators, where the conventional low-frequency picture of the resistive-wall effect is no longer sufficient and a more general treatment is required \cite{metralResistivewallImpedanceInfinitely2007}.

A major step in this direction was the development of a field-matching formalism for infinitely long cylindrical structures, which yields consistent expressions for the longitudinal and transverse impedances for arbitrary beam velocity, frequency, conductivity, permittivity, and permeability. This framework was first established for one- and two-layer cylindrical beam pipes and was later extended to an arbitrary number of surrounding layers, as well as to other geometries such as parallel plates \cite{metralResistivewallImpedanceInfinitely2007,MounetThesis,impedancewake2d}. These developments made it possible to describe a broad class of multilayer impedance problems in which the beam still propagates in vacuum, while the surrounding media are allowed to have general electromagnetic properties.

In the context of a future muon collider, however, an additional extension of the formalism is needed. Reaching the target luminosity requires ionization cooling of the muon beam, in which the beam passes through a sequence of material absorbers in order to reduce its transverse and longitudinal emittances \cite{MuonCollider2025,MICE}. In such a situation, the assumption that the beam propagates in vacuum is no longer valid. Instead, the electromagnetic response of the material occupying the beam region must be included directly in the impedance model.

The purpose of this paper is to derive analytical expressions for the beam-coupling impedance when the beam propagates in matter inside a cylindrically symmetric multilayer structure. A first treatment of impedance in matter with perfectly conducting surroundings was presented in Ref.~\cite{metral_hb2025}. Here, this approach is generalized to multilayer configurations with arbitrary surrounding materials. Relative to the vacuum-beam-region formalism, the extension requires two modifications: the source term in the scalar Helmholtz equation acquires a material-dependent factor, and the radial propagation constant is modified by the electromagnetic properties of the beam medium. Using the field-matching approach, we obtain the longitudinal and transverse impedances for general electromagnetic properties of both the beam region and the surrounding layers, which may be frequency dependent. The resulting formulation extends the conventional multilayer impedance framework to configurations relevant for material absorbers in muon-collider cooling channels.

In addition to the analytical derivation, we investigate the physical consequences of introducing material in the beam region. We first consider perfectly conducting surroundings in order to isolate the material response, examining how the material factor $F$ governs the frequency-dependent amplitude and sign of the impedance, while the radial propagation constant $\nu_1$ determines the radial field behavior and the appearance of resonant modes. The direct and indirect space-charge contributions are discussed separately, and the corresponding indirect space-charge wake functions are obtained numerically in the time domain. We then study how the impedance is modified by finite-conductivity surrounding layers. Finally, the applicability of the infinite-length approximation is discussed through comparison with existing finite-length mode-matching calculations for a vacuum beam region.

\section{Analytical formalism}

\begin{figure}
    \centering
    \includegraphics[width=0.8\linewidth]{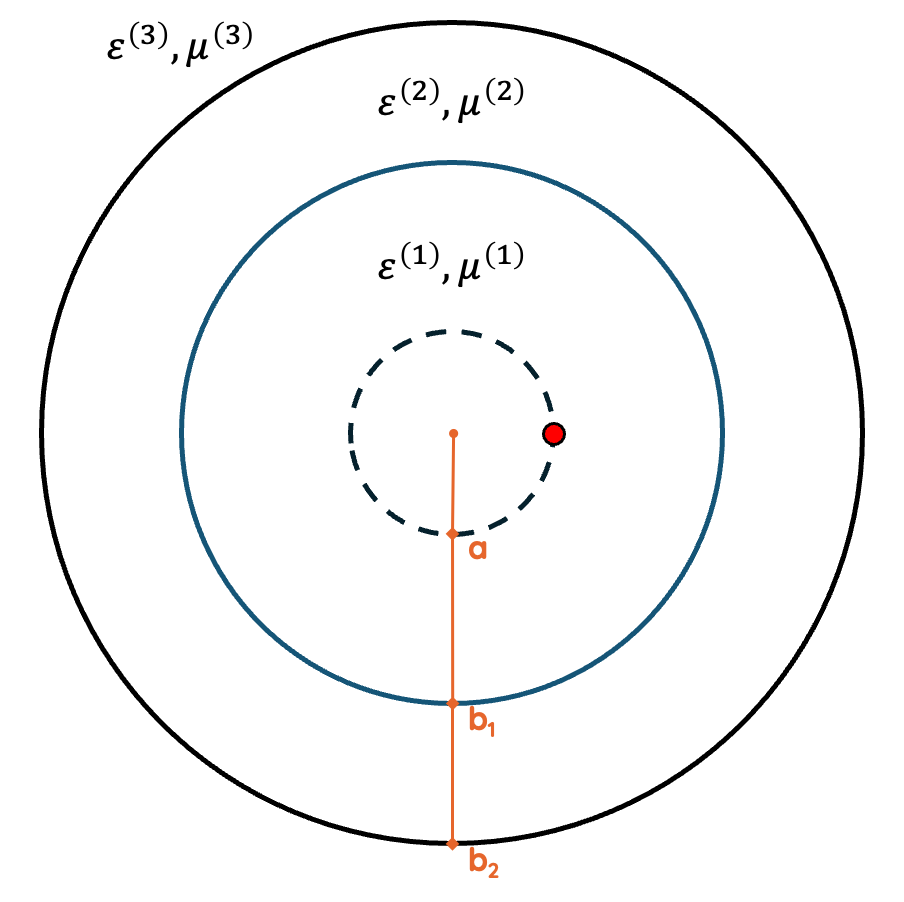}
    \caption{Schematic of the axisymmetric multilayer structure considered in this work. The beam propagates in region 1, which may be filled with material, while the surrounding cylindrical layers may also have arbitrary, frequency-dependent electromagnetic properties.}
    \label{fig:structure}
\end{figure}

\subsection{General formulation}

We follow the field-matching approach developed in Ref.~\cite{metralResistivewallImpedanceInfinitely2007} for cylindrically symmetric multilayer structures. In the present case, the formalism is extended from a vacuum beam region to a beam propagating through matter, while the surrounding regions are allowed to have arbitrary electromagnetic properties, as shown in Fig.~\ref{fig:structure}. In region $i$, the constitutive relations are written as
\begin{equation}
\varepsilon^{(i)}(\omega)=\varepsilon_0 \varepsilon_1^{(i)}(\omega),
\qquad
\mu^{(i)}(\omega)=\mu_0 \mu_1^{(i)}(\omega),
\end{equation}
where $\varepsilon_0$ and $\mu_0$ are the permittivity and permeability of free space, respectively. The quantities $\varepsilon_1^{(i)}(\omega)$ and $\mu_1^{(i)}(\omega)$ denote the relative permittivity and permeability of layer $i$ and may, in general, be complex and frequency dependent. Here, $\omega$ is the angular frequency. A commonly used representation is to incorporate the conductivity into the complex relative permittivity as

\begin{equation}
\varepsilon_1^{(i)}(\omega)
=
\varepsilon_r^{(i)}
+
\frac{\sigma^{(i)}}{j\varepsilon_0\omega},
\end{equation}
where $\varepsilon_r^{(i)}$ is the dielectric relative permittivity, $\sigma^{(i)}$ is the electrical conductivity of layer $i$, and $j$ is the imaginary unit. More general dispersive behavior can still be included through the complex functions $\varepsilon_1^{(i)}(\omega)$ and $\mu_1^{(i)}(\omega)$. In the calculations and physical discussion presented in this work, $\varepsilon_r^{(i)}$ and $\mu_1^{(i)}$ are taken to be real.

The beam travels along the $z$ axis with velocity $v=\beta c$, where $c$ is the speed of light in vacuum and $\beta$ is the normalized beam velocity. We use the time-harmonic convention $e^{j\omega t}$ and the longitudinal dependence $e^{-jkz}$, with $k=\omega/v$.

The beam source is modeled as a point-like charge moving parallel to the $z$ axis. In cylindrical coordinates $(r,\vartheta,z)$, the corresponding frequency-domain charge density can be expanded in azimuthal modes indexed by $m$ as
\begin{equation}
\rho_m(r,\vartheta,z;\omega)
=
\frac{Q}{av}\,\delta(r-a)\,e^{-jkz}\,
\frac{\cos(m\vartheta)}{\pi(1+\delta_{m0})},
\end{equation}
where $Q$ is the source charge, $a$ is the source radius, $\delta(r-a)$ is the Dirac delta function, and $\delta_{m0}$ is the Kronecker delta. The associated current density is
\begin{equation}
\mathbf{J}_m = \rho_m v\,\hat{\mathbf e}_z .
\end{equation}
In this work, the monopolar mode $m=0$ and the dipolar mode $m=1$ are considered, corresponding to the longitudinal and transverse impedances, respectively.

Substitution of the source into the $z$ component of the wave equation gives the source contribution
\begin{equation}
\frac{1}{\varepsilon}\frac{\partial \rho_m}{\partial z}
+
j\omega\mu\rho_m v
=
-j\rho_m\frac{k}{\varepsilon_0}F,
\end{equation}
where the material factor in the beam region is
\begin{equation}
\label{eq:F_main}
F \equiv \frac{1}{\varepsilon_1^{(1)}}-\mu_1^{(1)}\beta^2.
\end{equation}
Thus, relative to the vacuum case, the factor $1/\gamma^2$, where $\gamma=(1-\beta^2)^{-1/2}$, is replaced by $F$. The second modification is the radial propagation constant in region $i$,
\begin{equation}
\label{eq:nu_main}
\nu_i = k\sqrt{1-\beta^2 \varepsilon_1^{(i)}\mu_1^{(i)}},
\end{equation}
which replaces the usual vacuum expression $k/\gamma$ in the beam region. In vacuum, $\varepsilon_1^{(1)}=\mu_1^{(1)}=1$, so that $F=1-\beta^2=1/\gamma^2$ and $\nu_1=k/\gamma$.

Starting from Maxwell's equations, the longitudinal electric field in the beam region satisfies an inhomogeneous scalar Helmholtz equation. Away from the source term, the solutions in each homogeneous region are therefore linear combinations of modified Bessel functions. For azimuthal mode number $m$, the longitudinal field components can be written as
\begin{subequations}
\begin{align}
E_z^{(i)}(r,\vartheta,z;\omega) &=
\left[A_i I_m(\nu_i r)+B_i K_m(\nu_i r)\right]\cos(m\vartheta)e^{-jkz},
\\
H_z^{(i)}(r,\vartheta,z;\omega) &=
\left[C_i I_m(\nu_i r)+D_i K_m(\nu_i r)\right]\sin(m\vartheta)e^{-jkz},
\end{align}
\end{subequations}
where the coefficients are determined from continuity of the tangential fields at each material interface. Regularity at the origin and decay in the outermost region remove the singular or divergent solutions where appropriate. The detailed derivation of the electromagnetic fields used to obtain the impedance expressions is given in Appendix~\ref{app:derivation}.

\subsection{Longitudinal monopolar impedance}

The longitudinal monopolar impedance is obtained from the monopolar mode $m=0$. We define
\begin{equation}
x_0 \equiv \nu_1 a,
\qquad
x_1 \equiv \nu_1 b_1 .
\end{equation}
Following the standard derivation based on continuity of the longitudinal electric field at $r=a$, the total longitudinal impedance can be written as

\begin{equation}
\label{eq:Zlong_final_main}
Z_{\parallel}^{\mathrm{tot}}(\omega)
=
-\,j\,\frac{L \omega \mu_0}{2\pi \beta^2}I_0^2(x_0)
\left[
\frac{K_0(x_0)}{I_0(x_0)}-\alpha_{\mathrm{TM}}^{(0)}
\right]F,
\end{equation}
where $L$ is the length of the structure and $\alpha_{\mathrm{TM}}^{(0)}$ is a dimensionless reflection coefficient associated with the monopolar TM mode. The superscript denotes the azimuthal mode number.

Equation~\eqref{eq:Zlong_final_main} has the same structure as in the vacuum case, but with the substitutions $1/\gamma^2 \to F$ and $k/\gamma \to \nu_1$. The first term in brackets corresponds to the direct space-charge contribution, while the coefficient $\alpha_{\mathrm{TM}}^{(0)}$ contains the response of the surrounding layers. In the special case of a perfect electric conductor (PEC) located at radius $b_1$, one simply has
\begin{equation}\label{eq:alpha_tm_long}
\alpha_{\mathrm{TM}}^{(0)}=\frac{K_0(x_1)}{I_0(x_1)},
\end{equation}
and Eq.~\eqref{eq:Zlong_final_main} reduces to the direct-plus-indirect space-charge result for a material-filled beam region surrounded by a perfect conductor \cite{metral_hb2025}.

For general multilayer surroundings, $\alpha_{\mathrm{TM}}^{(0)}$ is obtained from the continuity of $E_z$ and $H_\vartheta$ at each interface. The analytical expressions remain compact for a single external layer, but become increasingly lengthy for multiple layers. In the present work, these coefficients are obtained systematically from the field-matching system and evaluated numerically when needed.

For a single external layer extending to infinity outside the beam region, introducing $x_2 \equiv \nu_2 b_1$, the monopolar reflection coefficient is
\begin{equation}
    \alpha_{\mathrm{TM}}^{(0)}
    =
    \frac{
    \varepsilon_1^{(1)}\nu_{2}K_0(x_2)K_0'(x_1)
    -
    \varepsilon_1^{(2)}\nu_{1}K_0(x_1)K_0'(x_2)
    }{
    \varepsilon_1^{(1)}\nu_{2}I_0'(x_1)K_0(x_2)
    -
    \varepsilon_1^{(2)}\nu_{1}I_0(x_1)K_0'(x_2)
    }.
    \label{eq:alphaTM0_onelayer}
\end{equation}
For a vacuum beam region, $\varepsilon_1^{(1)}=\mu_1^{(1)}=1$ and $\nu_1=k/\gamma$, and Eq.~\eqref{eq:alphaTM0_onelayer} reduces to the expression of Ref.~\cite{metralResistivewallImpedanceInfinitely2007}.

\subsection{Transverse dipolar impedance}

The transverse dipolar impedance is obtained from the dipolar mode $m=1$. Following the same field-matching procedure as in the longitudinal case, the total horizontal impedance can be written as

\begin{equation}
\label{eq:Ztrans_final_main}
Z_x^{\mathrm{tot}}(\omega)=
-\,j\,L\,\frac{Z_0}{\pi\beta}\,
\frac{I_1^2(x_0)}{a^2}
\left[
\frac{K_1(x_0)}{I_1(x_0)}-\alpha_{\mathrm{TM}}^{(1)}
\right]F,
\end{equation}
where $Z_0=\sqrt{\mu_0/\varepsilon_0}$ and $\alpha_{\mathrm{TM}}^{(1)}$ is the reflection coefficient associated with the dipolar TM mode. As in the longitudinal case, the term $K_1(x_0)/I_1(x_0)$ represents the direct space-charge contribution, whereas $\alpha_{\mathrm{TM}}^{(1)}$ contains the effect of the surrounding structure. For a perfect conductor at $r=b_1$,
\begin{equation}\label{eq:alpha_tm_trans}
\alpha_{\mathrm{TM}}^{(1)}=\frac{K_1(x_1)}{I_1(x_1)},
\end{equation}

\noindent and Eq.~\eqref{eq:Ztrans_final_main} reduces to the direct-plus-indirect space-charge result for a material-filled beam region \cite{metral_hb2025}. For general material layers, the determination of $\alpha_{\mathrm{TM}}^{(1)}$ requires matching the dipolar field components at each interface. An auxiliary coefficient associated with the TE-like part of the field appears in the intermediate algebra, but the final transverse impedance depends only on $\alpha_{\mathrm{TM}}^{(1)}$.

For a single external layer extending to infinity, it is convenient to introduce
\begin{equation}
\begin{aligned}
    P_1 &\equiv \frac{I_1'(x_1)}{I_1(x_1)},
    &
    Q_1 &\equiv \frac{K_1'(x_1)}{K_1(x_1)},
    &
    Q_2 &\equiv \frac{K_1'(x_2)}{K_1(x_2)}.
\end{aligned}
\end{equation}
We further define
\begin{subequations}
\begin{align}
    A &\equiv \nu_{2}\mu_1^{(1)}P_1-\nu_{1}\mu_1^{(2)}Q_2, 
    \label{eq:alphaTM1_A}
    \\
    B &\equiv \nu_{2}\varepsilon_1^{(1)}P_1-\nu_{1}\varepsilon_1^{(2)}Q_2, \\
    C &\equiv \nu_{2}x_2-\nu_{1}x_1,
\end{align}
\begin{equation}
    \alpha_{\mathrm{TM}}^{(1)}
    =
    \frac{K_1(x_1)}{I_1(x_1)}
    \left[
    1+
    \frac{
        \nu_2\varepsilon_1^{(1)}(P_1-Q_1)(\beta x_1 x_2)^2 A
    }{
        C^2-(\beta x_1 x_2)^2AB
    }
    \right].
    \label{eq:alphaTM1_onelayer}
\end{equation}
\end{subequations}
For a vacuum beam region, $\varepsilon_1^{(1)}=\mu_1^{(1)}=1$ and $\nu_1=k/\gamma$, and Eq.~\eqref{eq:alphaTM1_onelayer} reduces to the expression of Ref.~\cite{metralResistivewallImpedanceInfinitely2007}.

\subsection{Determination of the interface coefficients}

The coefficients $\alpha_{\mathrm{TM}}^{(0)}$ and $\alpha_{\mathrm{TM}}^{(1)}$ encode the response of the surrounding layers. For the monopolar problem, they are obtained by imposing continuity of $E_z$ and $H_\vartheta$ at each cylindrical interface. For the dipolar problem, continuity of $E_z$, $G_z$, $E_\vartheta$, and $G_\vartheta$ is imposed, where $\mathbf{G}=Z_0\mathbf{H}$.

For a single external layer, the resulting reflection coefficients are given by Eq.~\eqref{eq:alphaTM0_onelayer} and Eqs.~\eqref{eq:alphaTM1_A}--\eqref{eq:alphaTM1_onelayer}. For multiple external layers, the same field-matching procedure leads to a linear system for the unknown field amplitudes, which is solved numerically. The analytical expressions and interface-matching relations were implemented in an in-house code to evaluate the longitudinal and transverse impedances for a material-filled beam region surrounded by up to two external cylindrical layers.

\section{Impedance for absorber-relevant configurations}

The final cooling stage of a future muon collider is designed to reduce the transverse emittance of the muon bunch while allowing the longitudinal emittance to increase. This is achieved by passing the beam through low-$Z$ absorber material, such as liquid hydrogen (LH$_2$) or lithium hydride (LiH), in strong solenoidal focusing fields. In the design considered in Ref.~\cite{MuonCollider2025}, the transverse emittance is reduced from approximately $140\,\mu\mathrm{m}$ to $22.5\,\mu\mathrm{m}$.

\begin{figure}
    \centering
    \includegraphics[width=\linewidth]{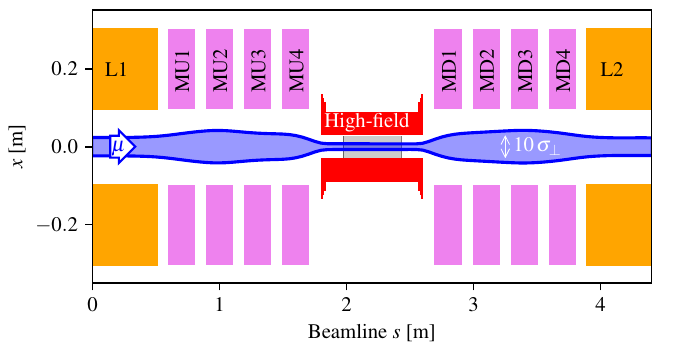}
    \caption{Example lattice of a final-cooling cell for a future muon collider. The $10\sigma$ beam envelope is shown in blue, while the low-field solenoids and matching coils are shown in orange and pink, respectively. The absorber is shown in gray and is surrounded by the high-field solenoid in red \cite{Stechauner_IPAC26_coolpy}.}
    \label{fig:final_cooling_schematic}
\end{figure}

The formalism derived above assumes an infinitely long cylindrical structure. It is therefore expected to be most applicable when the longitudinal extent of the absorber is larger than the characteristic transverse scale, here represented by the first material-interface radius $b_1$. In the final-cooling configurations considered here, the absorber radius is approximately $b_1=25\,\mathrm{mm}$, while the absorber length lies in the range $L=500$--$1200\,\mathrm{mm}$ \cite{stechaunerPrivateCommunication2026}. The corresponding ratio is therefore $L/b_1=20$--$48$, so the absorber length is much larger than its radius. An example final-cooling cell is shown in Fig.~\ref{fig:final_cooling_schematic}. This makes the infinite-length approximation a good first model for this regime, as further discussed in Sec.~\ref{subsec:finite_length}.

\subsection{Reference configuration}
\label{subsec:ref_config}

The calculations presented below use a common reference geometry defined by $\beta=0.5$, $a=0.1\,\mathrm{mm}$, $b_1=1\,\mathrm{cm}$, and $L=1\,\mathrm{m}$. The electromagnetic properties of the beam region and of the surrounding material are varied separately in the following subsections in order to distinguish their respective contributions to the impedance.

\subsection{Material response with perfectly conducting surroundings}
\label{subsec:pec_material_response}

To isolate the effects associated with the material filling the beam region from those introduced by the surrounding layers, we first consider the limiting case of a perfectly conducting boundary at $r=b_1$. In this case, $\alpha_{\mathrm{TM}}^{(m)}$ is given by Eqs.~\eqref{eq:alpha_tm_long} and \eqref{eq:alpha_tm_trans} for the longitudinal and transverse impedances, respectively. The resulting impedance contains the direct and indirect space-charge contributions, and is denoted in the following by $Z_{\parallel,x}^{\mathrm{SC}}$. The beam-region material enters through two quantities: the factor $F$, which multiplies the source term and sets an overall complex amplitude, and the radial propagation constant $\nu_1$, which governs the radial field dependence through the Bessel functions and, in particular, the appearance of resonant modes.

\subsubsection{Material factor and total space-charge response}

As a reference, for a vacuum beam region the material factor reduces to $F=\gamma^{-2}$. For the material model considered below, $\varepsilon_r^{(1)}$, $\mu_1^{(1)}$, and $\sigma^{(1)}$ are taken to be real and frequency independent. Substitution of the corresponding complex permittivity into Eq.~\eqref{eq:F_main} gives
\begin{equation}
\label{eq:F_expanded}
\begin{aligned}
F ={}&
\frac{\varepsilon_r^{(1)}}
{\left(\varepsilon_r^{(1)}\right)^2
+\left(\sigma^{(1)}/\varepsilon_0\omega\right)^2}
-\mu_1^{(1)}\beta^2
\\
&+j
\frac{\sigma^{(1)}/\varepsilon_0\omega}
{\left(\varepsilon_r^{(1)}\right)^2
+\left(\sigma^{(1)}/\varepsilon_0\omega\right)^2}.
\end{aligned}
\end{equation}

\noindent The material factor is therefore in general complex. For $\sigma^{(1)}>0$, its real part varies from $-\mu_1^{(1)}\beta^2$ in the low-frequency limit to $1/\varepsilon_r^{(1)}-\mu_1^{(1)}\beta^2$ at high frequency. Its imaginary part vanishes in both limits and reaches a maximum at $\omega=\sigma^{(1)}/(\varepsilon_0\varepsilon_r^{(1)})$. Thus, for fixed $\varepsilon_r^{(1)}$ and $\mu_1^{(1)}$, changing the conductivity shifts the characteristic material-response frequency while leaving the limiting values of $\mathrm{Re}(F)$ unchanged.

The same limiting values can also be understood by considering the conductivity at fixed frequency. For $\sigma^{(1)}\rightarrow 0$, the material factor approaches $1/\varepsilon_r^{(1)}-\mu_1^{(1)}\beta^2$, whereas for $\sigma^{(1)}\rightarrow\infty$ the electric contribution $1/\varepsilon_1^{(1)}$ vanishes and $F\rightarrow-\mu_1^{(1)}\beta^2$. The latter limit corresponds to strong screening of the electric field in a highly conducting beam-region material, leaving only the magnetic contribution to the material factor.

\begin{figure}
    \centering
    \includegraphics[width=\linewidth]{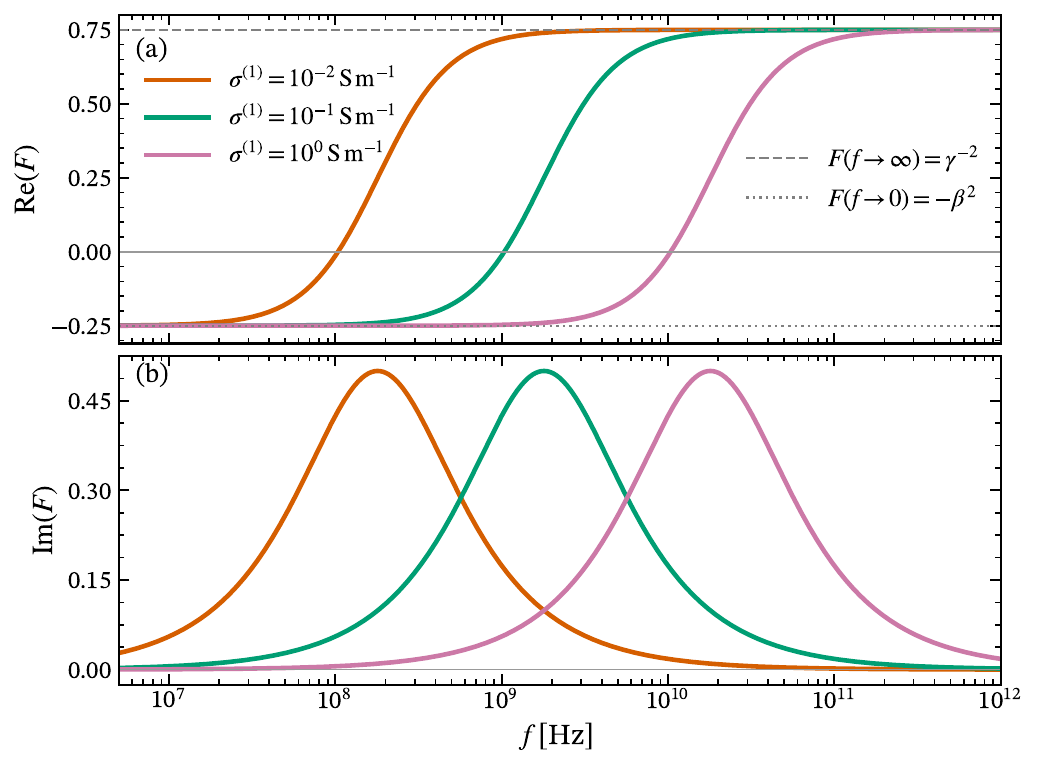}
    \caption{Real (a) and imaginary (b) parts of the material factor $F$ as a function of frequency for different conductivities $\sigma^{(1)}$, with $\varepsilon_r^{(1)}=\mu_1^{(1)}=1$ and $\beta=0.5$. The dashed and dotted horizontal lines indicate the high- and low-frequency limits, $F\rightarrow\gamma^{-2}$ and $F\rightarrow-\beta^2$, respectively.}
    \label{fig:F_conductivity_scan}
\end{figure}

This behavior is illustrated in Fig.~\ref{fig:F_conductivity_scan} for $\varepsilon_r^{(1)}=\mu_1^{(1)}=1$ and $\beta=0.5$. In this case, $\mathrm{Re}(F)$ changes sign between its low-frequency value $-\beta^2$ and its high-frequency limit $\gamma^{-2}$, which coincides with the vacuum value. Since the impedance expressions contain an overall factor $-jF$, this sign change can reverse the sign of the imaginary impedance, while a finite $\mathrm{Im}(F)$ gives rise to a real component. Increasing $\sigma^{(1)}$ shifts these features towards higher frequencies.

\begin{figure}
    \centering
    \includegraphics[width=\linewidth]{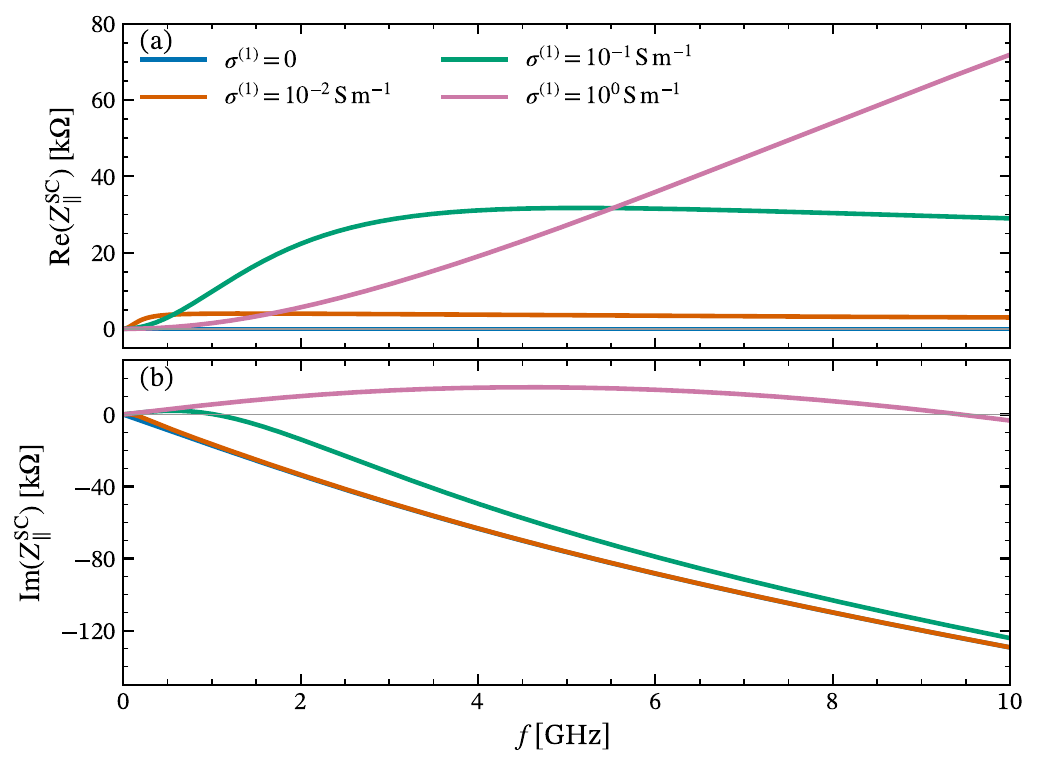}
    \caption{Real (a) and imaginary (b) parts of the total longitudinal space-charge impedance for a beam region with $\varepsilon_r^{(1)}=\mu_1^{(1)}=1$ and different conductivities $\sigma^{(1)}$, surrounded by a perfectly conducting boundary. The vacuum result, $\sigma^{(1)}=0$, is included for comparison.}
    \label{fig:sc_impedance_long}
\end{figure}
\begin{figure}
    \centering
    \includegraphics[width=\linewidth]{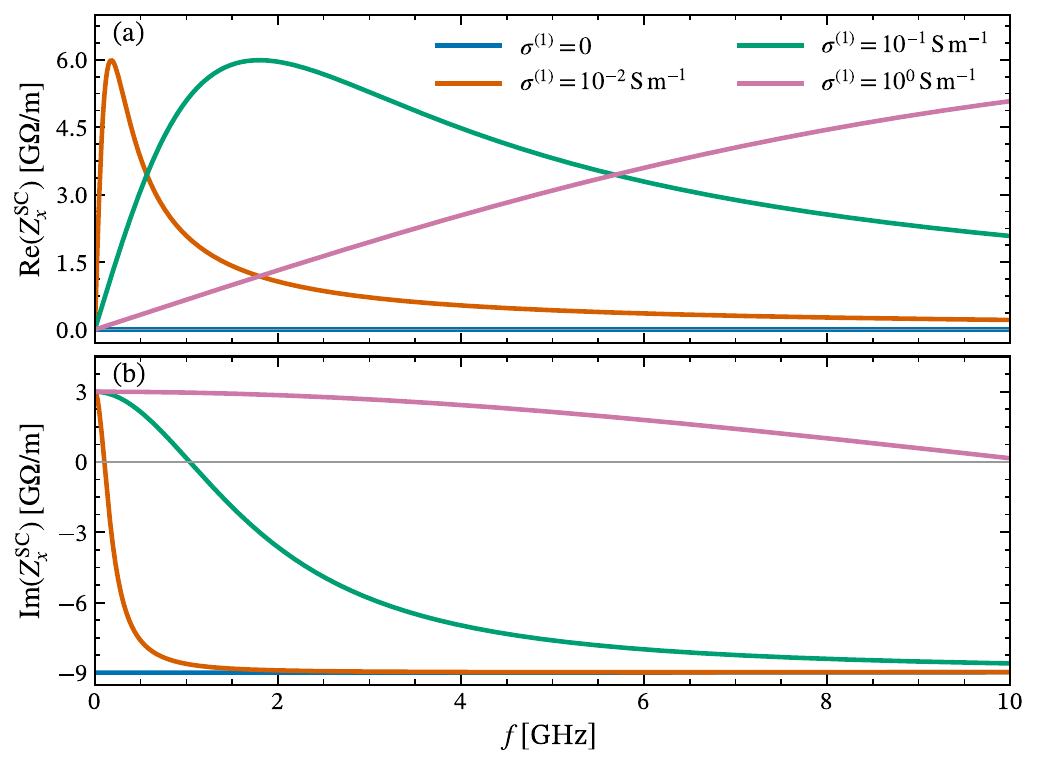}
    \caption{Real (a) and imaginary (b) parts of the total transverse space-charge impedance for the same configuration as in Fig.~\ref{fig:sc_impedance_long}. The vacuum result, $\sigma^{(1)}=0$, is included for comparison.}
    \label{fig:sc_impedance_trans}
\end{figure}

The corresponding changes in the total space-charge impedance are shown in Figs.~\ref{fig:sc_impedance_long} and \ref{fig:sc_impedance_trans}. In the small-argument regime, $|\nu_1 b_1|\ll1$, the perfectly conducting expressions reduce to
\begin{equation}
\label{eq:SC_small_argument}
\begin{aligned}
Z_{\parallel}^{\mathrm{SC}}
&\simeq
-j\frac{L\omega\mu_0}{2\pi\beta^2}
\ln\left(\frac{b_1}{a}\right)F ,
\\[3pt]
Z_x^{\mathrm{SC}}
&\simeq
-j\frac{LZ_0}{2\pi\beta}
\left(
\frac{1}{a^2}-\frac{1}{b_1^2}
\right)F .
\end{aligned}
\end{equation}
\noindent These expressions make the role of the material factor particularly transparent. In the transverse plane, the impedance is directly proportional to $-jF$, whereas the longitudinal impedance contains an additional explicit factor of $\omega$. A finite $\mathrm{Im}(F)$ therefore generates a real impedance in both planes, while a sign change of $\mathrm{Re}(F)$ reverses the sign of the imaginary component. The additional factor of $\omega$ suppresses the corresponding low-frequency features in the longitudinal plane, whereas they appear directly in the transverse impedance. Increasing $\sigma^{(1)}$ shifts these material-response features towards higher frequencies.

For the reference geometry, the total transverse impedance in Fig.~\ref{fig:sc_impedance_trans} is strongly dominated by the direct space-charge contribution, and its frequency dependence therefore closely follows that of $F$. The maximum of the real part consequently occurs near the maximum of $\mathrm{Im}(F)$, while the sign change of $\mathrm{Re}(F)$ produces a reversal of the imaginary impedance. In the convention used here, the negative imaginary vacuum direct-space-charge impedance corresponds to the usual defocusing space-charge force; the reversal of the imaginary impedance therefore corresponds to a reversal of the transverse force towards a focusing response. The large magnitude of the transverse impedance follows from the strong $1/a^2$ dependence in Eq.~\eqref{eq:SC_small_argument} for $a\ll b_1$.

\subsubsection{Direct and indirect space charge}

It is useful to separate the perfectly conducting result into its direct and indirect space-charge contributions,
\begin{equation}
\label{eq:SC_decomposition}
Z_{\parallel,x}^{\mathrm{SC}}
=
Z_{\parallel,x}^{\mathrm{DSC}}
+
Z_{\parallel,x}^{\mathrm{ISC}} .
\end{equation}
\noindent The direct contribution corresponds to the self-field of the source in an unbounded beam-region medium, whereas the indirect contribution describes the modification of this field by the surrounding boundary. This distinction is also useful in beam-dynamics calculations, where the direct space-charge force depends on the transverse beam distribution and is commonly evaluated self-consistently, while the boundary-induced contribution can be represented through an impedance or wake function. In the present point-source formulation, the separation is particularly relevant for the transverse impedance, for which the direct contribution becomes very large as the source radius $a$ is reduced.

\noindent In the small-argument regime $|\nu_1 b_1|\ll1$, the transverse direct and indirect contributions reduce to
\begin{equation}
\label{eq:trans_dsc_isc_small}
\begin{aligned}
Z_x^{\mathrm{DSC}}
&\simeq
-j\frac{LZ_0}{2\pi\beta a^2}F ,
\\[3pt]
Z_x^{\mathrm{ISC}}
&\simeq
+j\frac{LZ_0}{2\pi\beta b_1^2}F .
\end{aligned}
\end{equation}
\noindent Their magnitude ratio therefore scales as $(b_1/a)^2$. For the reference configuration, $b_1/a=100$, so the transverse total space-charge impedance is dominated by the direct contribution by approximately four orders of magnitude in this limit. The indirect contribution is therefore shown separately below. Besides isolating the response associated with the surrounding boundary, this permits comparison with established indirect-space-charge impedance and wake-function results without the strong dependence of the direct point-source contribution on $a$.

\begin{figure}
    \centering
    \includegraphics[width=\linewidth]{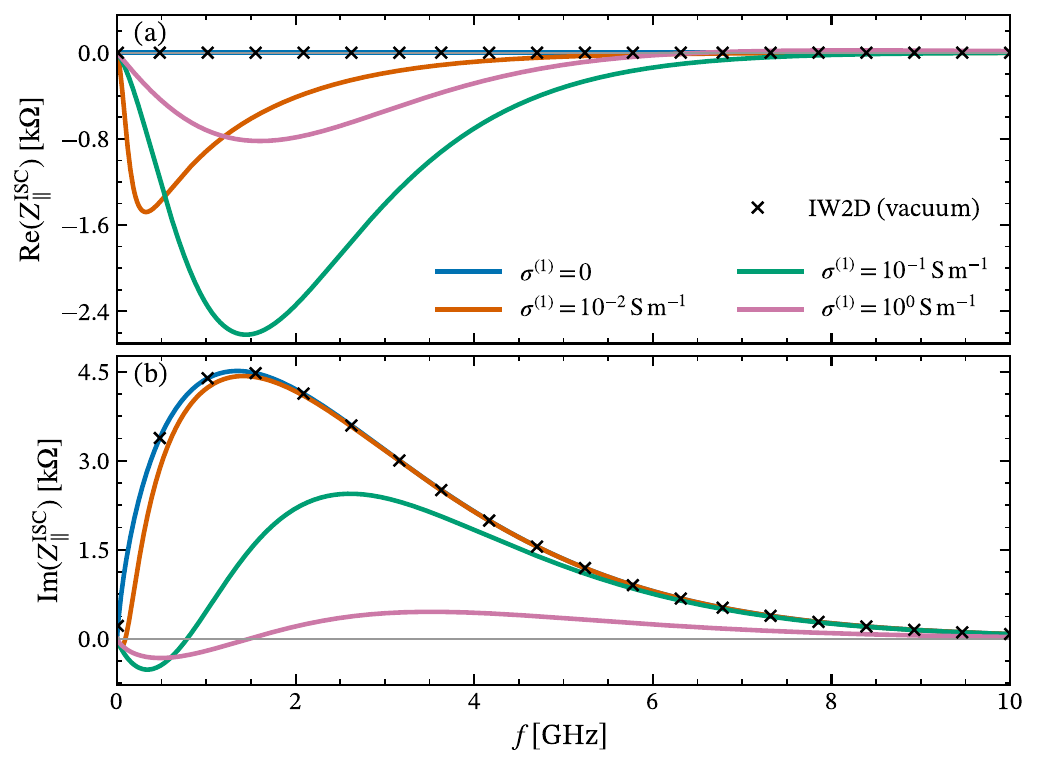}
    \caption{Real (a) and imaginary (b) parts of the longitudinal indirect space-charge impedance for a beam region with $\varepsilon_r^{(1)}=\mu_1^{(1)}=1$ and different conductivities $\sigma^{(1)}$, surrounded by a perfectly conducting boundary. The vacuum result, $\sigma^{(1)}=0$, is included for comparison. Crosses indicate the corresponding IW2D~\cite{impedancewake2d} results for the vacuum case.}
    \label{fig:isc_impedance_long}
\end{figure}

\begin{figure}
    \centering
    \includegraphics[width=\linewidth]{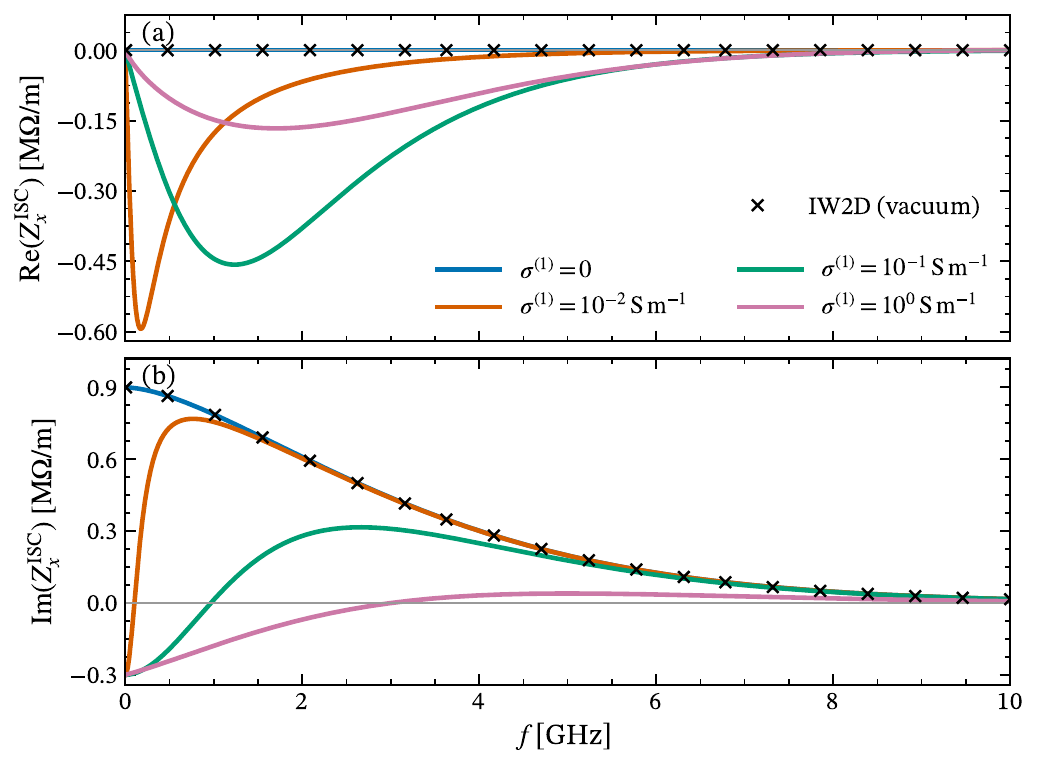}
    \caption{Real (a) and imaginary (b) parts of the transverse indirect space-charge impedance for the same configuration as in Fig.~\ref{fig:isc_impedance_long}. The vacuum result, $\sigma^{(1)}=0$, is included for comparison. Crosses indicate the corresponding IW2D~\cite{impedancewake2d} results for the vacuum case.}
    \label{fig:isc_impedance_trans}
\end{figure}

The corresponding longitudinal and transverse indirect space-charge impedances are shown in Figs.~\ref{fig:isc_impedance_long} and \ref{fig:isc_impedance_trans}. In the vacuum limit, the indirect contribution has the opposite sign to the direct contribution for perfectly conducting surroundings. For a material-filled beam region, however, the indirect impedance does not in general follow the frequency dependence of $F$ alone. The boundary response also depends on the radial propagation constant through modified Bessel functions evaluated at $\nu_1 b_1$. Since $\nu_1$ is generally complex for a conductive beam-region material, these factors acquire both magnitude and phase variations. Consequently, the maxima of the real impedance and the zero crossings of its imaginary part need not coincide with the corresponding features of $F$. The simple proportionality to $F$ in Eq.~\eqref{eq:trans_dsc_isc_small} is recovered only when $|\nu_1 b_1|\ll1$. This departure is more pronounced for the indirect contribution than for the direct contribution because the relevant radial scale is $b_1$ rather than the much smaller source radius $a$.

To validate the vacuum limit of the present formulation, the results for $\sigma^{(1)}=0$ in Figs.~\ref{fig:isc_impedance_long} and \ref{fig:isc_impedance_trans} are compared with independent calculations using IW2D, based on the multilayer impedance formalism~\cite{MounetThesis}. Since IW2D computes the wall impedance, which reduces to the indirect space-charge impedance in the PEC limit, a surrounding layer with conductivity $\sigma^{(2)}=10^{15}\,\mathrm{S\,m^{-1}}$ is used to approximate the perfectly conducting boundary. The results show agreement for both the longitudinal and transverse impedances. \rev{An additional comparison with IW2D for a finite-conductivity multilayer LHC collimator is presented in Appendix~\ref{app:iw2d_validation}.}

\subsubsection{Indirect space-charge wake functions}

To examine the boundary-induced response in the time domain, we introduce the longitudinal wake coordinate $s=z_{\mathrm{s}}-z_{\mathrm{w}}$, where $z_{\mathrm{s}}$ and $z_{\mathrm{w}}$ denote the positions of the source and witness particles, respectively. The beam propagates in the positive $z$ direction, such that $s>0$ corresponds to a witness trailing the source. Using the impedance convention adopted in this work, the longitudinal and transverse wake functions are related to the corresponding impedances by
%

\begin{subequations}
\label{eq:wake_transform}
\begin{align}
W_{\parallel}(s)
&=
\frac{1}{2\pi}
\int_{-\infty}^{\infty}
Z_{\parallel}(\omega)
e^{j\omega s/v}\,d\omega ,
\\
W_x(s)
&=
-\frac{j}{2\pi}
\int_{-\infty}^{\infty}
Z_x(\omega)
e^{j\omega s/v}\,d\omega .
\end{align}
\end{subequations}

\begin{figure}
    \centering
    \includegraphics[width=\linewidth]{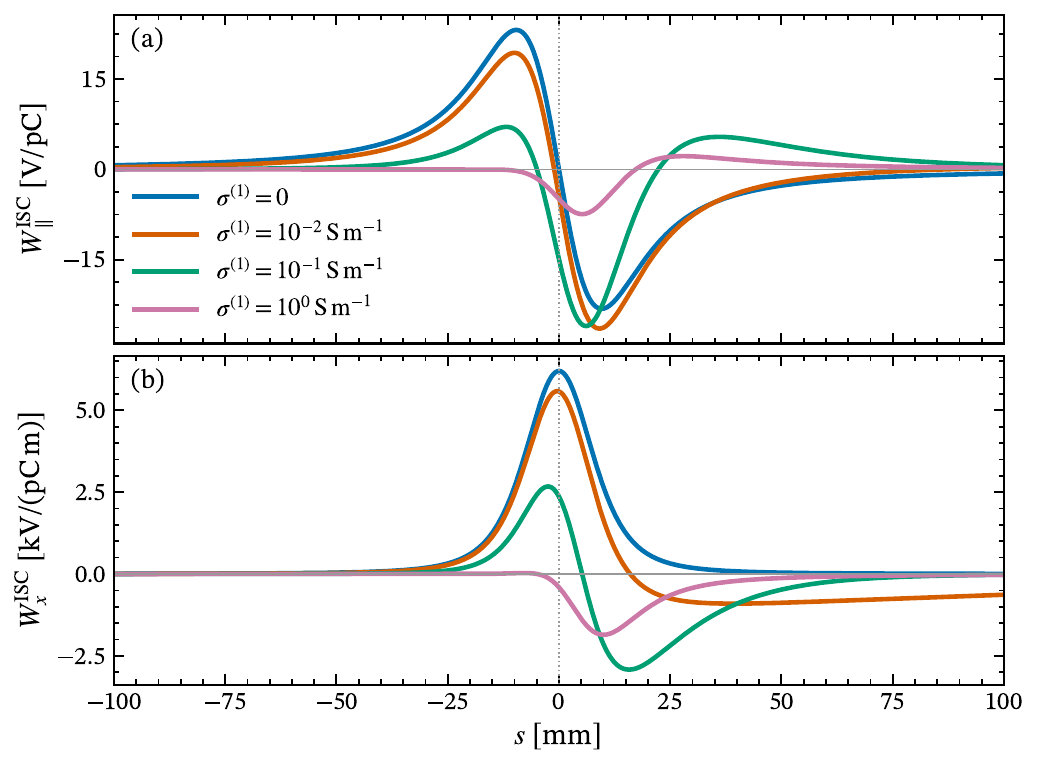}
    \caption{Longitudinal (a) and transverse (b) indirect space-charge wake functions corresponding to the impedances in Figs.~\ref{fig:isc_impedance_long} and \ref{fig:isc_impedance_trans}. Positive $s$ corresponds to a witness trailing the source.}
    \label{fig:wake_function_isc}
\end{figure}

The inverse transforms are evaluated numerically on a discrete frequency grid using an inverse fast Fourier transform. In the following, we apply these transforms to the indirect space-charge impedances of Figs.~\ref{fig:isc_impedance_long} and \ref{fig:isc_impedance_trans}, thereby isolating the time-domain response associated with the surrounding boundary. The resulting wake functions are shown in Fig.~\ref{fig:wake_function_isc}. Increasing the beam-region conductivity modifies both the magnitude and shape of the wake. In the vacuum case, both ISC impedances are purely imaginary. The longitudinal impedance is odd in frequency and therefore gives an antisymmetric longitudinal wake about $s=0$, whereas the transverse impedance is even in frequency and gives a symmetric transverse wake. A finite conductivity introduces real impedance components, which modify these symmetries. In the transverse plane, the real component adds an antisymmetric contribution to the wake and shifts its extrema away from the source position. In the longitudinal plane, it adds a symmetric contribution to the otherwise antisymmetric vacuum wake. For the parameters considered here, increasing $\sigma^{(1)}$ also reduces the wake magnitude, and at the largest conductivity shown the transverse wake reverses polarity relative to the vacuum result.

As a validation of the inverse-transform procedure, the vacuum indirect-space-charge wake functions were benchmarked against the closed-form expressions of Ref.~\cite{mounetClosedFormFormulas2022} and found to agree.


\subsubsection{Radial propagation and resonances}

We now turn to the resonant behavior associated with the radial propagation constant $\nu_1$. Such resonances can appear in beam-region materials with sufficiently large permittivity. For a lossless material, $\nu_1$ is real when $\beta^2\varepsilon_r^{(1)}\mu_1^{(1)}<1$, corresponding to an evanescent radial dependence described by modified Bessel functions. This includes the vacuum case for any $\beta<1$. When $\beta^2\varepsilon_r^{(1)}\mu_1^{(1)}>1$, however, $\nu_1$ becomes imaginary. Writing $\nu_1=jk_\perp$, the identity $I_m(jk_\perp r)=j^mJ_m(k_\perp r)$ shows that the radial dependence can instead be expressed in terms of the ordinary Bessel function $J_m$ and becomes oscillatory. The condition $\beta^2\varepsilon_r^{(1)}\mu_1^{(1)}>1$ is equivalent to the beam velocity exceeding the electromagnetic phase velocity $c/\sqrt{\varepsilon_r^{(1)}\mu_1^{(1)}}$ in the material.

For a perfectly conducting boundary at $r=b_1$, the indirect space-charge term contains the boundary factor $K_m(\nu_1b_1)/I_m(\nu_1b_1)$ and therefore develops poles when $I_m(\nu_1b_1)=0$. In the lossless oscillatory regime, this reduces to the resonance condition $J_m(k_\perp b_1)=0$. The corresponding resonance frequencies are

\begin{equation}
\label{eq:resonance_frequencies}
f_n =
\frac{\beta c}
{2\pi b_1
\sqrt{\beta^2\varepsilon_r^{(1)}\mu_1^{(1)}-1}}
j_{m,n},
\end{equation}
\noindent where $j_{m,n}$ is the $n$th positive zero of the ordinary Bessel function $J_m$. The longitudinal monopole impedance corresponds to $m=0$, while the transverse dipole impedance corresponds to $m=1$. Equation~\eqref{eq:resonance_frequencies} shows directly that the resonance frequencies scale as $1/b_1$ and decrease as $\varepsilon_r^{(1)}\mu_1^{(1)}$ increases. For large $n$, the spacing between successive zeros of $J_m$ approaches $\pi$ \cite{abramowitzHandbookMathematicalFunctions1965}, so that the higher-order resonances become approximately equally spaced.

\begin{figure}
    \centering
    \includegraphics[width=\linewidth]{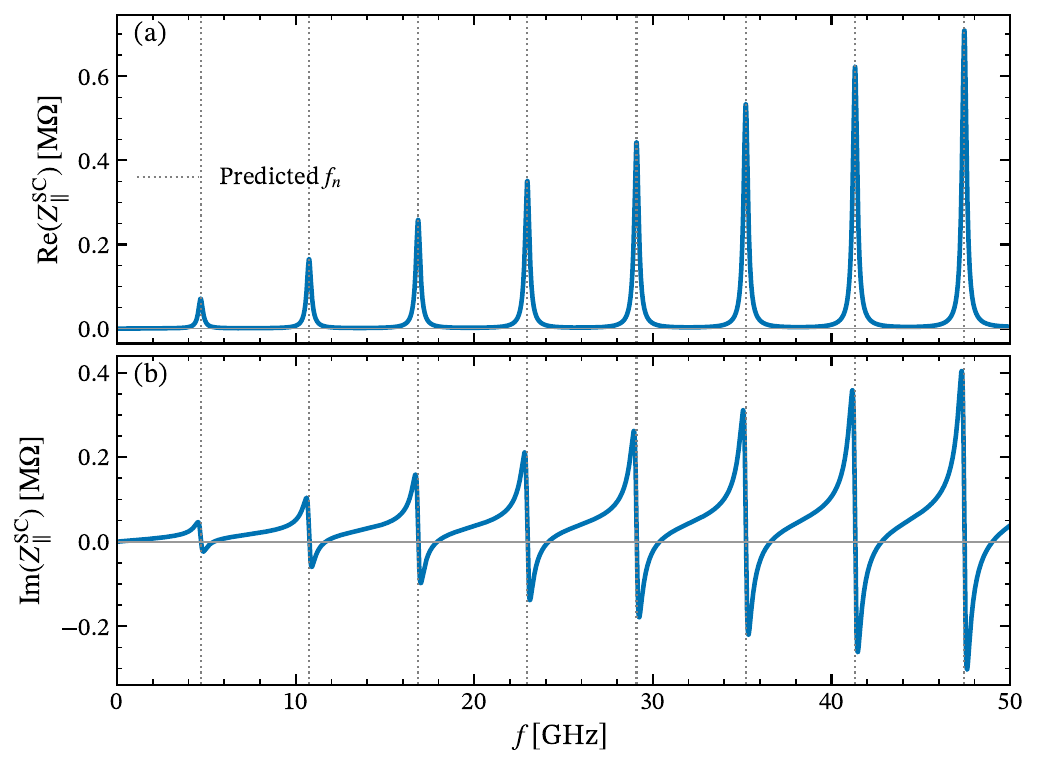}
    \caption{Real (a) and imaginary (b) parts of the total longitudinal space-charge impedance for a material-filled beam region with $\varepsilon_r^{(1)}=10$, $\mu_1^{(1)}=1$, and $\sigma^{(1)}=10^{-1}~\mathrm{S\,m^{-1}}$, surrounded by a perfectly conducting boundary at $b_1=1~\mathrm{cm}$. The vertical dotted lines indicate the lossless monopole resonance frequencies predicted by Eq.~\eqref{eq:resonance_frequencies} with $m=0$.}
    \label{fig:resonances_long}
\end{figure}

\begin{figure}
    \centering
    \includegraphics[width=\linewidth]{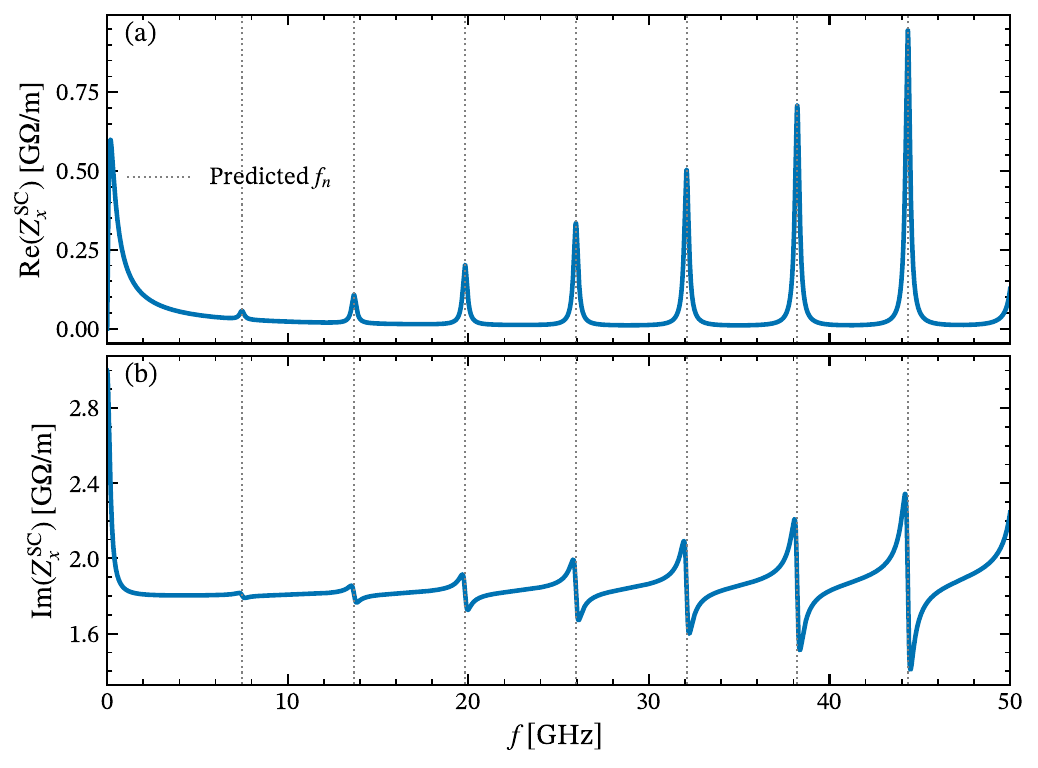}
    \caption{Real (a) and imaginary (b) parts of the total transverse space-charge impedance for the same configuration as in Fig.~\ref{fig:resonances_long}. The vertical dotted lines indicate the lossless dipole resonance frequencies predicted by Eq.~\eqref{eq:resonance_frequencies} with $m=1$.}
    \label{fig:resonances_trans}
\end{figure}

Figures~\ref{fig:resonances_long} and \ref{fig:resonances_trans} show the longitudinal and transverse total space-charge impedances for $\varepsilon_r^{(1)}=10$, $\mu_1^{(1)}=1$, $\beta=0.5$, and $b_1=1~\mathrm{cm}$. A finite conductivity $\sigma^{(1)}=10^{-1}~\mathrm{S\,m^{-1}}$ is included, which gives the resonances a finite width. The vertical dotted lines show the lossless resonance frequencies predicted by Eq.~\eqref{eq:resonance_frequencies}. Finite conductivity makes $\nu_1$ complex, so that the corresponding poles are displaced from the real-frequency axis. For the parameters considered here, this primarily broadens the resonances, while the shift of their peak frequencies remains small, explaining the close agreement with the lossless prediction.

The transverse result in Fig.~\ref{fig:resonances_trans} also contains a smooth positive imaginary background and a low-frequency maximum in its real part. These are nonresonant features associated with the material response through $F$, rather than additional modes. For the parameters considered here, the high-frequency limit $1/\varepsilon_r^{(1)}-\mu_1^{(1)}\beta^2<0$ gives a positive imaginary background through the factor $-jF$, while the low-frequency real feature is associated with $\mathrm{Im}(F)$. The same material response is present longitudinally, but the additional factor of $\omega$ suppresses the low-frequency feature relative to the resonant structures.

\subsection{Effect of finite-conductivity surroundings}

We next consider how the impedance is modified when the perfectly conducting boundary is replaced by an external medium of finite conductivity. Since the direct space-charge contribution is determined entirely by the beam-region material and is unaffected by the surroundings, it is useful to isolate the boundary-induced contribution. Following Ref.~\cite{roncaroloComparisonLaboratoryMeasurements2009}, we define the wall impedance as

\begin{equation}
\label{eq:wall_impedance}
Z_{\parallel,x}^{\mathrm{wall}}
=
Z_{\parallel,x}^{\mathrm{tot}}
-
Z_{\parallel,x}^{\mathrm{DSC}} .
\end{equation}
\noindent The wall impedance includes both the contribution that remains for perfectly conducting surroundings, i.e. the indirect space-charge impedance, and the additional contribution arising from the finite conductivity and, more generally, the electromagnetic properties of the surrounding material. Thus, in the PEC limit, $Z_{\parallel,x}^{\mathrm{wall}}=Z_{\parallel,x}^{\mathrm{ISC}}$.

The reference geometry of Sec.~\ref{subsec:ref_config} is retained. To isolate the effect of the external conductivity, the beam-region conductivity is fixed to $\sigma^{(1)}=10^{-1}\,\mathrm{S\,m^{-1}}$ and the permeability to $\mu_1^{(1)}=1$ in both configurations. We consider a unity-permittivity configuration with $\varepsilon_r^{(1)}=1$ and a high-permittivity configuration with $\varepsilon_r^{(1)}=10$. The external medium is assigned the same relative permittivity and permeability as the corresponding beam region, $\varepsilon_r^{(2)}=\varepsilon_r^{(1)}$ and $\mu_1^{(2)}=\mu_1^{(1)}=1$, while its conductivity is varied as $\sigma^{(2)}=10^{-1},\,10^0,\,10^2\,\mathrm{S\,m^{-1}}$ for the unity-permittivity configuration and $\sigma^{(2)}=10^{-1},\,10^2,\,10^4\,\mathrm{S\,m^{-1}}$ for the high-permittivity configuration.

\begin{figure}[t]
    \centering
    \includegraphics[width=\linewidth]{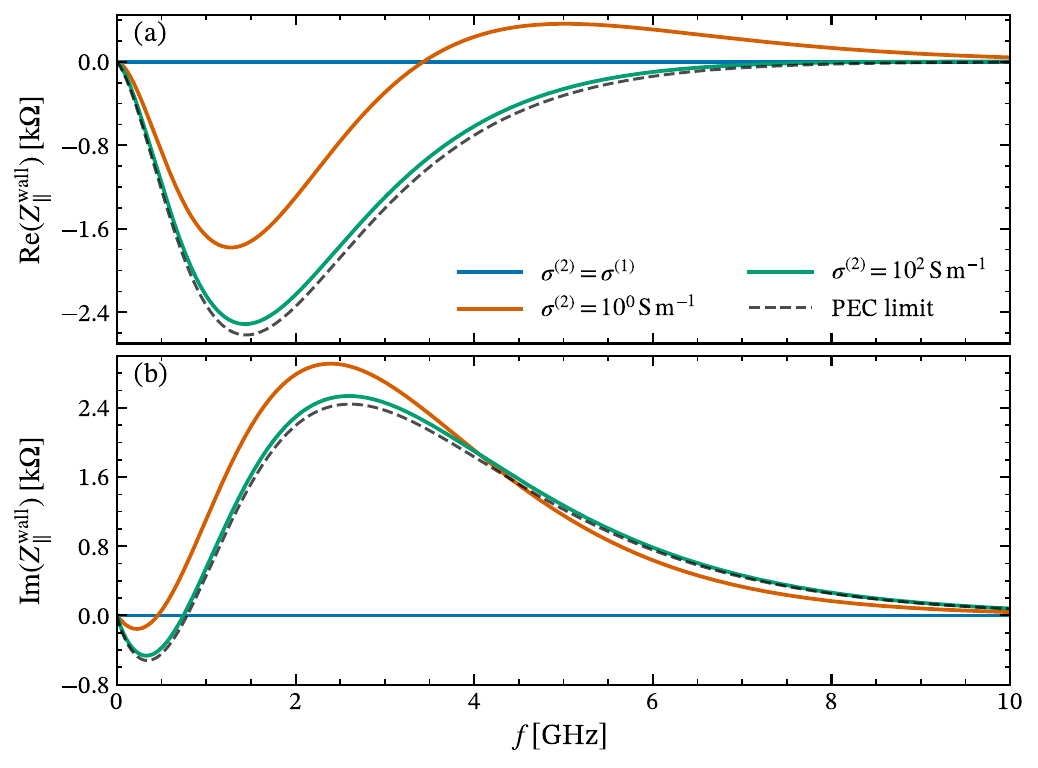}
    \caption{Real (a) and imaginary (b) parts of the longitudinal wall impedance $Z_{\parallel}^{\mathrm{wall}}$ as a function of frequency for the unity-permittivity configuration, with $\varepsilon_r^{(1)}=\varepsilon_r^{(2)}=1$, $\mu_1^{(1)}=\mu_1^{(2)}=1$, and $\sigma^{(1)}=10^{-1}\,\mathrm{S\,m^{-1}}$. The colored curves correspond to different external conductivities $\sigma^{(2)}$, while the dashed black curve shows the PEC limit, for which $Z_{\parallel}^{\mathrm{wall}}=Z_{\parallel}^{\mathrm{ISC}}$.}
    \label{fig:Zlong_A}
\end{figure}

\begin{figure}[t]
    \centering
    \includegraphics[width=\linewidth]{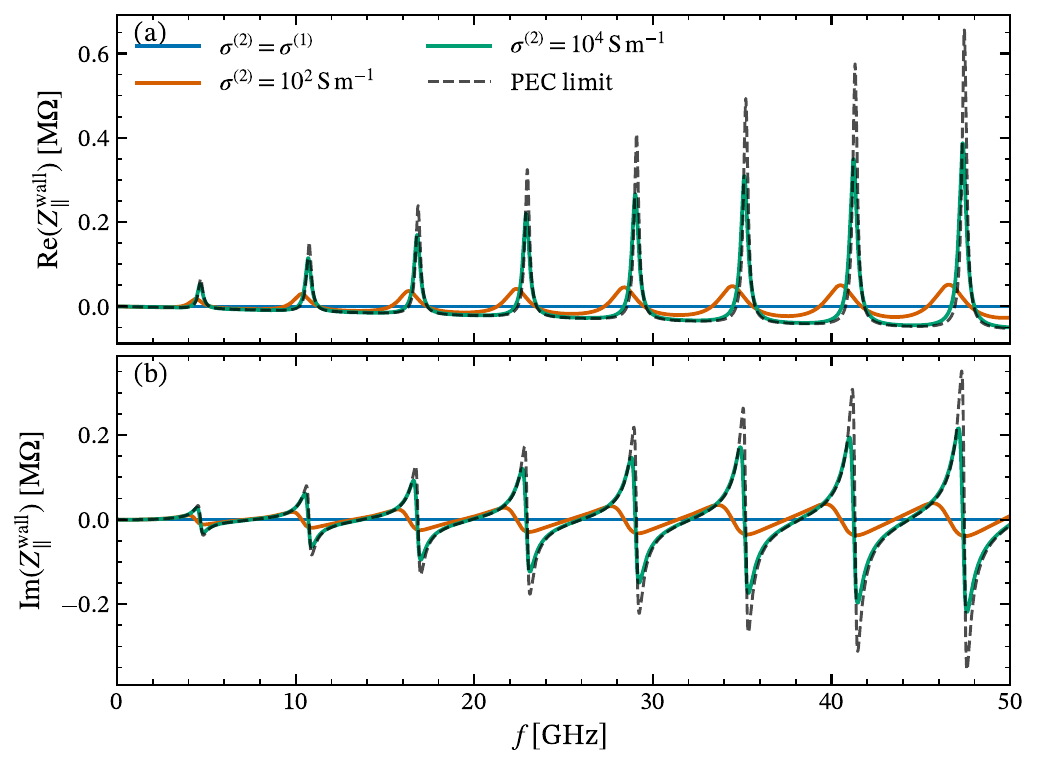}
    \caption{Same as Fig.~\ref{fig:Zlong_A}, but for the high-permittivity configuration, with $\varepsilon_r^{(1)}=\varepsilon_r^{(2)}=10$ and external conductivities $\sigma^{(2)}=10^{-1},\,10^2,\,10^4\, \mathrm{S\,m^{-1}}$.}
    \label{fig:Zlong_B}
\end{figure}

\begin{figure}[t]
    \centering
    \includegraphics[width=\linewidth]{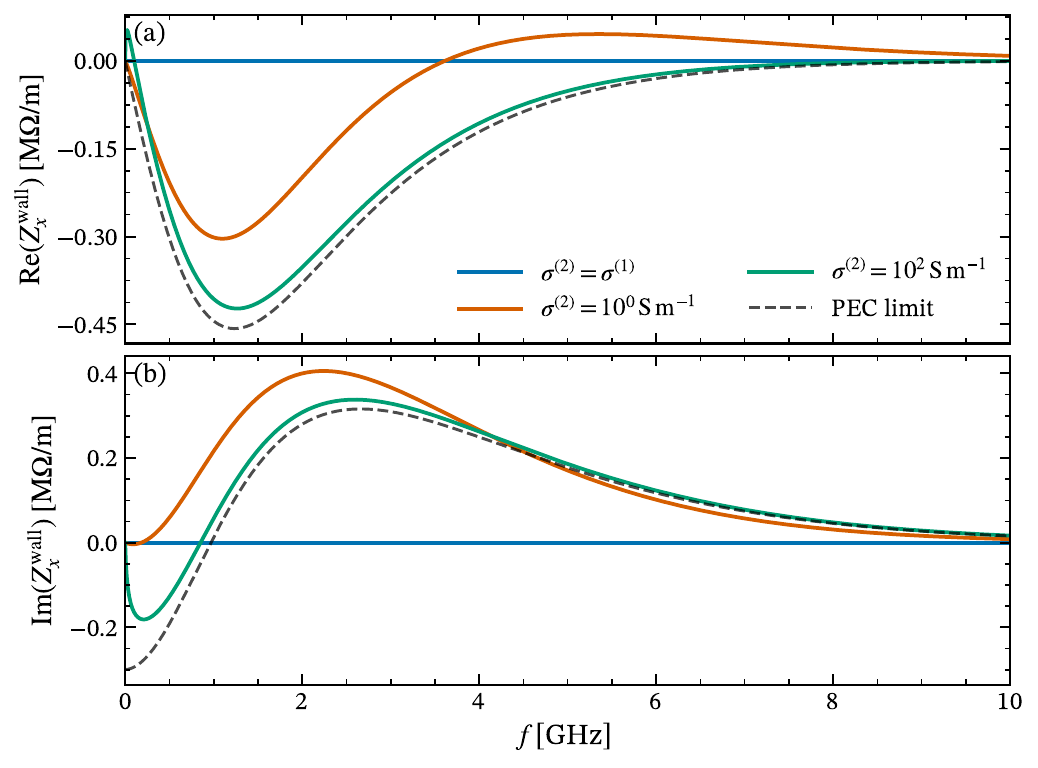}
    \caption{Real (a) and imaginary (b) parts of the transverse wall impedance $Z_x^{\mathrm{wall}}$ as a function of frequency for the unity-permittivity configuration. The colored curves correspond to different external conductivities $\sigma^{(2)}$, while the dashed black curve shows the PEC limit, for which $Z_x^{\mathrm{wall}}=Z_x^{\mathrm{ISC}}$.}
    \label{fig:Ztrans_A}
\end{figure}

\begin{figure}[t]
    \centering
    \includegraphics[width=\linewidth]{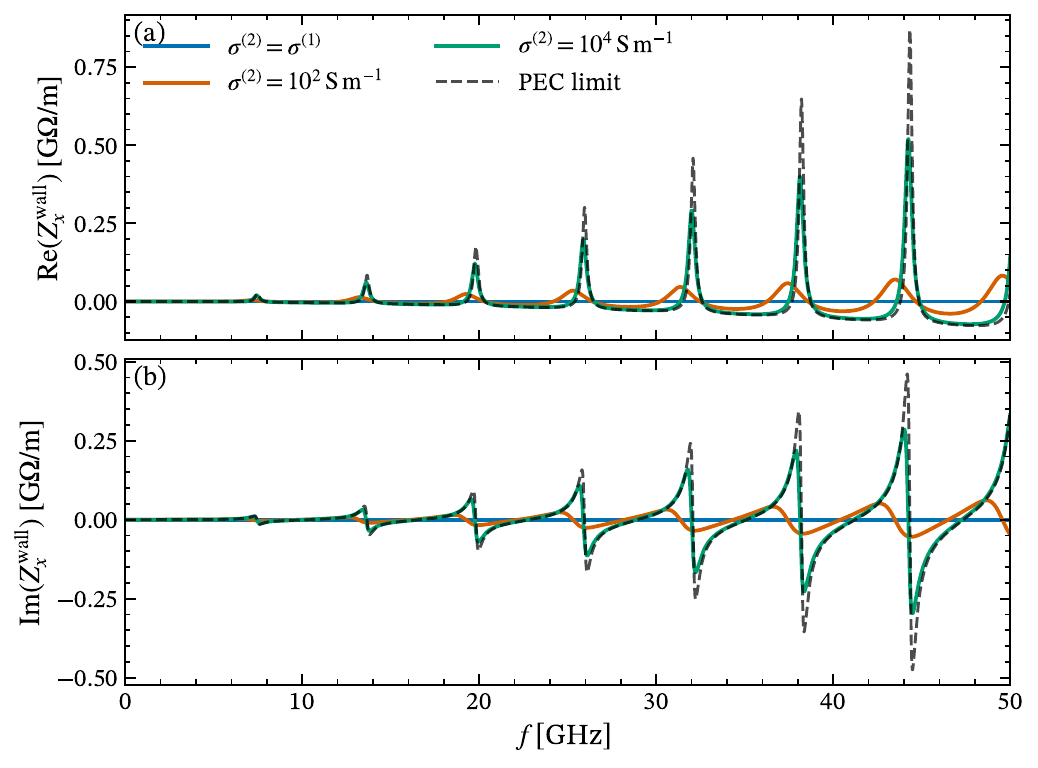}
    \caption{Same as Fig.~\ref{fig:Ztrans_A}, but for the high-permittivity configuration, with $\varepsilon_r^{(1)}=\varepsilon_r^{(2)}=10$ and external conductivities $\sigma^{(2)}=10^{-1},\,10^2,\,10^4\, \mathrm{S\,m^{-1}}$.}
    \label{fig:Ztrans_B}
\end{figure}

Figures~\ref{fig:Zlong_A}--\ref{fig:Ztrans_B} show the corresponding longitudinal and transverse wall impedances. For $\sigma^{(2)}=\sigma^{(1)}$, all material parameters are identical on either side of the interface, such that there is no discontinuity in the material properties and $Z_{\parallel,x}^{\mathrm{wall}}=0$. As the conductivity of the external medium is increased, the wall impedance approaches the corresponding PEC indirect-space-charge result. This convergence occurs more rapidly for the unity-permittivity configuration than for the high-permittivity configuration. For a given conductivity and frequency, the ratio of the magnitudes of the conduction and displacement currents scales as $\sigma^{(2)}/(\omega\varepsilon_0\varepsilon_r^{(2)})$. The external medium with $\varepsilon_r^{(2)}=10$ therefore retains a stronger dielectric response than that with $\varepsilon_r^{(2)}=1$ at the same $\sigma^{(2)}$. In the high-permittivity configuration, the approach towards the PEC limit is also reflected in the resonant structure, with the resonant peaks becoming increasingly pronounced as $\sigma^{(2)}$ is increased.

\subsection{Remarks on finite-length effects}\label{subsec:finite_length}

The formalism developed in this paper assumes an infinitely long cylindrical structure and is expected to be most accurate when $L>b_1$, where $L$ is the length of the structure and $b_1$ is the radius of the first material interface.

For some absorber configurations in the ionization cooling channel, this condition is not strictly satisfied. In particular, during the rectilinear cooling stages, the on-axis length of the LiH absorbers can be as small as $6.5\,\mathrm{cm}$, comparable to a transverse radius of approximately $8\,\mathrm{cm}$~\cite{MuonCollider2025}. In such cases, finite-length effects may play a non-negligible role, and the applicability of the infinite-length approximation should be assessed carefully.

\begin{figure}
    \centering
    \includegraphics[width=1\linewidth]{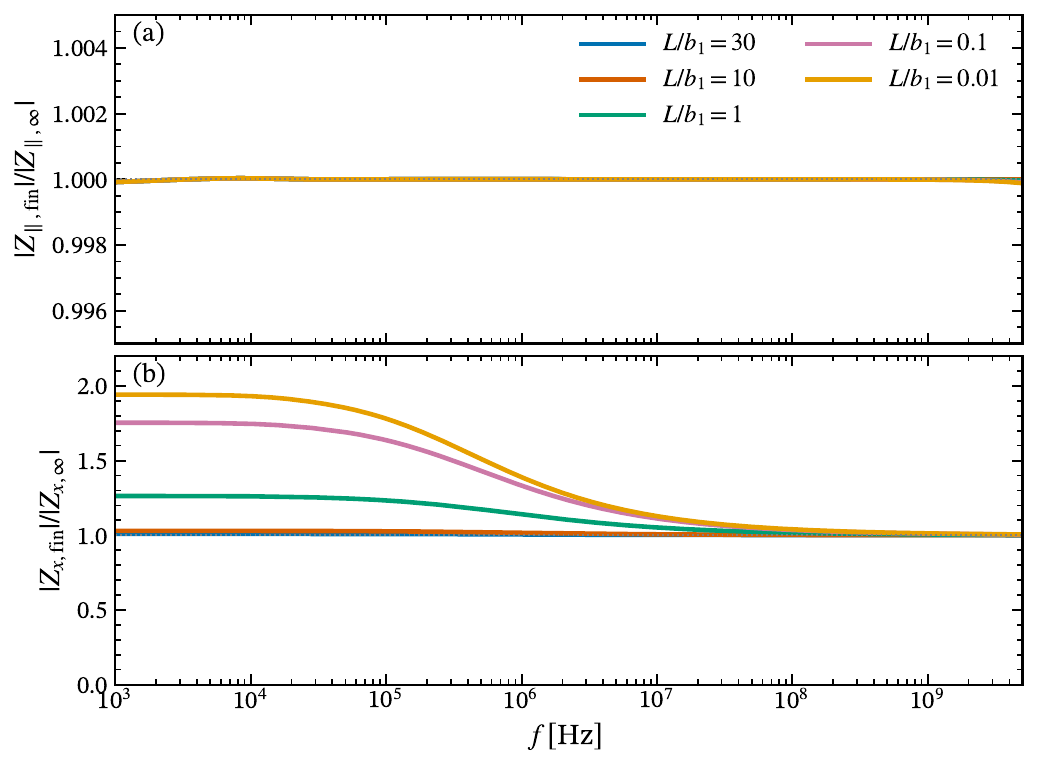}
    \caption{Ratio between the absolute values of the finite-length and infinite-length impedance models as a function of frequency for different ratios $L/b_1$. The longitudinal ratio $|Z_{\parallel,\mathrm{fin}}|/|Z_{\parallel,\infty}|$ is shown in (a), and the transverse ratio $|Z_{x,\mathrm{fin}}|/|Z_{x,\infty}|$ is shown in (b). The finite-length results are obtained using the mode-matching methods of Refs.~\cite{Biancacci2014,Biancacci2023}.}
    \label{fig:finite_length}
\end{figure}

To investigate the range of validity of the infinite-length approximation, we compare the present infinite-length expressions with finite-length mode-matching calculations developed in Refs.~\cite{Biancacci2014,Biancacci2023}. Since these finite-length methods are formulated for a vacuum beam region, the comparison is performed for a vacuum beam pipe rather than for the material-filled beam-region case. The geometry is defined by $b_1=2\,\mathrm{mm}$ and $b_2=2.5\,\mathrm{cm}$, with the second material characterized by $\varepsilon_r^{(2)}=1$, $\mu_1^{(2)}=1$, and $\sigma^{(2)}=10^{5}\,\mathrm{S\,m^{-1}}$. The outer boundary is treated as a perfect electric conductor.

The length $L$ of the finite structure is varied relative to the inner radius $b_1$, and the resulting impedances are compared with the corresponding infinite-length expressions. In the following, $Z_{\parallel,\mathrm{fin}}$ and $Z_{x,\mathrm{fin}}$ denote the longitudinal and transverse impedances obtained from the finite-length mode-matching calculations, while $Z_{\parallel,\infty}$ and $Z_{x,\infty}$ denote the corresponding quantities obtained from the infinite-length model. Figure~\ref{fig:finite_length} shows the absolute-value ratios $|Z_{\parallel,\mathrm{fin}}|/|Z_{\parallel,\infty}|$ and $|Z_{x,\mathrm{fin}}|/|Z_{x,\infty}|$.

Figure~\ref{fig:finite_length}(a) shows that the longitudinal ratio remains close to unity over the frequency range considered, indicating good agreement between the finite- and infinite-length models for this configuration. For the transverse impedance, shown in Fig.~\ref{fig:finite_length}(b), the ratio is more sensitive to the value of $L/b_1$. At low frequencies, the ratio approaches an approximately frequency-independent value whose magnitude varies with $L/b_1$, while at higher frequencies it gradually approaches unity. The departure from the infinite-length result becomes larger as the structure is shortened, suggesting that finite-length effects may be particularly important for the transverse impedance of short absorber elements.

A useful frequency scale for interpreting the transverse comparison follows from the known frequency dependence of the impedance of a conducting cylindrical layer. Distinct low-, intermediate-, and high-frequency regimes can be identified \cite{metralTransverseResistivewallImpedance2005,roncaroloComparisonLaboratoryMeasurements2009}. For the present comparison, the relevant frequency scale is associated with the transition from the low-frequency inductive-bypass regime to the intermediate-frequency regime in which the classical thick-wall approximation applies. For a thick conducting layer, this transition occurs when the skin depth becomes comparable to the beam-pipe radius. In particular, the broad maximum of the real transverse impedance occurs approximately for $\delta_{\mathrm{skin}}\simeq b_1$ \cite{roncaroloComparisonLaboratoryMeasurements2009}. With $\delta_{\mathrm{skin}}=\sqrt{2/(\omega\mu^{(2)}\sigma^{(2)})}$, this gives the characteristic frequency
\begin{equation}
f_{\max,\mathrm{Re}}
\simeq
\frac{1}{\pi\mu^{(2)}\sigma^{(2)}b_1^2}.
\label{eq:fmax_re}
\end{equation}
\noindent For the parameters considered here, this yields $f_{\max,\mathrm{Re}}\simeq0.63~\mathrm{MHz}$. This frequency lies within the range over which the transverse finite-to-infinite impedance ratio in Fig.~\ref{fig:finite_length}(b) departs from its low-frequency plateau and begins to approach unity.

Because this comparison is performed for a vacuum beam region, it should be regarded only as an indication of the possible importance of finite-length effects. Dedicated electromagnetic simulations, for example with WAKIS \cite{de_la_Fuente_Garcia_Wakis_2025}, are required to assess finite-length effects for material-filled absorber geometries.

\section{Conclusion}

Analytical expressions for the monopolar longitudinal and dipolar transverse beam-coupling impedances have been derived for a material-filled beam region in a cylindrically symmetric multilayer structure. Relative to the conventional vacuum formalism, the extension is obtained through the replacement of the vacuum factor $1/\gamma^2$ by the material-dependent factor $F$ and of the vacuum radial propagation constant by its material-dependent counterpart.

Representative impedance computations were presented for parameter sets motivated by absorber configurations relevant to ionization cooling. For perfectly conducting surroundings, the material response was interpreted in terms of $F$ and the radial propagation constant $\nu_1$. The direct and indirect space-charge contributions were considered separately. A finite beam-region conductivity can generate a real impedance component and reverse the sign of the imaginary space-charge impedance. Beam-region materials with sufficiently high permittivity can give rise to resonant modes when the radial field becomes oscillatory, with resonance frequencies well described by the corresponding lossless Bessel-function condition. In the appropriate limit, the formalism recovers the conventional vacuum results. The effect of finite-conductivity surroundings was also examined. The possible importance of finite-length corrections was assessed through comparison with existing finite-length mode-matching results for a vacuum beam region. Dedicated electromagnetic simulations remain an important next step for material-filled absorber geometries.
\begin{acknowledgments}

The authors thank Xavier Buffat and Jos\'ephine Potdevin for fruitful discussions and for their pioneering work on this topic. This work is endorsed by the International Muon Collider Collaboration (IMCC).

\end{acknowledgments}

\appendix

\section{Derivation of the impedance expressions}
\label{app:derivation}

This appendix gives the main steps leading to Eqs.~\eqref{eq:Zlong_final_main} and \eqref{eq:Ztrans_final_main}. The derivation follows the multilayer cylindrical field-matching formalism of Ref.~\cite{metralResistivewallImpedanceInfinitely2007}, but extends it to the case where the beam region itself may have arbitrary electromagnetic properties. Additional details on the source-field construction and analytical impedance formalism can be found in Refs.~\cite{glucksternAnalyticMethodsCalculating2000,zotterNewResultsImpedance2005}. We use the same geometry and notation as in Sec.~II and Fig.~\ref{fig:structure}.

We use the time dependence $e^{j\omega t}$ and the longitudinal dependence $e^{-jkz}$, with
\begin{equation}
    k=\frac{\omega}{v}=\frac{\omega}{\beta c}.
\end{equation}
In region $i$, the constitutive relations are
\begin{subequations}
\begin{align}
    \mathbf{D}^{(i)}
    &=
    \varepsilon^{(i)}\mathbf{E}^{(i)}
    =
    \varepsilon_0\varepsilon_1^{(i)}\mathbf{E}^{(i)},
    \\
    \mathbf{B}^{(i)}
    &=
    \mu^{(i)}\mathbf{H}^{(i)}
    =
    \mu_0\mu_1^{(i)}\mathbf{H}^{(i)} .
\end{align}
\end{subequations}
Here $\varepsilon_1^{(i)}$ and $\mu_1^{(i)}$ may be complex and frequency dependent. When no layer index is shown, $\varepsilon$ and $\mu$ denote the absolute permittivity and permeability of the homogeneous region considered.

\subsection{Source term and azimuthal harmonics}

A point-like macroparticle of charge $Q$ moving parallel to the $z$ axis with velocity $v=\beta c$ and transverse displacement $a$ can be written as
\begin{equation}
    \rho(r,\vartheta,z;t)
    =
    \frac{Q}{a}\delta(r-a)\delta(\vartheta)\delta(z-vt).
\end{equation}
Using the cosine expansion
\begin{equation}
    \delta(\vartheta)
    =
    \sum_{m=0}^{\infty}
    \frac{\cos(m\vartheta)}{\pi(1+\delta_{m0})},
\end{equation}
the frequency-domain modal components become
\begin{equation}
    \rho_m(r,\vartheta,z;\omega)
    =
    \frac{Q}{av}\,\delta(r-a)\,e^{-jkz}\,
    \frac{\cos(m\vartheta)}{\pi(1+\delta_{m0})}.
    \label{eq:app_rho_m}
\end{equation}
The corresponding current density is
\begin{equation}
    \mathbf{J}_m=\rho_m v\,\hat{\mathbf e}_z .
\end{equation}

\subsection{Scalar equation for the longitudinal electric field}

Starting from Maxwell's equations,
\begin{subequations}
\label{eq:app_maxwell}
\begin{align}
    \nabla\cdot\mathbf{D} &= \rho, \\
    \nabla\cdot\mathbf{B} &= 0, \\
    \nabla\times\mathbf{E} &= -j\omega\mathbf{B},
    \label{eq:app_faraday}\\
    \nabla\times\mathbf{H} &= \mathbf{J}+j\omega\mathbf{D},
    \label{eq:app_ampere}
\end{align}
\end{subequations}
we take the curl of Eq.~\eqref{eq:app_faraday}. For a fixed frequency, $\varepsilon$ and $\mu$ may be complex, but they are spatially uniform within each homogeneous layer. Therefore,
\begin{equation}
    \nabla\times(\nabla\times\mathbf{E})
    =
    -j\omega\mu\left(\mathbf{J}+j\omega\varepsilon\mathbf{E}\right).
\end{equation}
Using
\begin{equation}
    \nabla\times(\nabla\times\mathbf{E})
    =
    \nabla(\nabla\cdot\mathbf{E})-\nabla^2\mathbf{E},
    \qquad
    \nabla\cdot\mathbf{E}=\frac{\rho}{\varepsilon},
\end{equation}
one obtains the inhomogeneous vector wave equation
\begin{equation}
    \left(\nabla^2+\omega^2\mu\varepsilon\right)\mathbf{E}
    =
    \nabla\left(\frac{\rho}{\varepsilon}\right)
    +
    j\omega\mu\mathbf{J}.
    \label{eq:app_vector_wave}
\end{equation}

We now write the $z$ component explicitly in cylindrical coordinates. This gives
\begin{widetext}
\begin{equation}
    \left[
    \frac{1}{r}\frac{\partial}{\partial r}
    \left(r\frac{\partial}{\partial r}\right)
    +
    \frac{1}{r^2}\frac{\partial^2}{\partial \vartheta^2}
    +
    \frac{\partial^2}{\partial z^2}
    +
    \omega^2\mu\varepsilon
    \right]E_z
    =
    \frac{\partial}{\partial z}\left(\frac{\rho}{\varepsilon}\right)
    +
    j\omega\mu J_z .
    \label{eq:app_wave_Ez}
\end{equation}
\end{widetext}
In the beam region, $J_z=\rho v$ and $\rho\propto e^{-jkz}$. The source term on the right-hand side then becomes
\begin{equation}
    \frac{1}{\varepsilon}\frac{\partial\rho}{\partial z}
    +
    j\omega\mu\rho v
    =
    -j\rho\frac{k}{\varepsilon_0}F ,
\end{equation}
where
\begin{equation}
    F\equiv
    \frac{1}{\varepsilon_1^{(1)}}-\mu_1^{(1)}\beta^2 .
    \label{eq:app_F}
\end{equation}
Thus the usual vacuum factor $1/\gamma^2$ is replaced by the material factor $F$. In vacuum, $\varepsilon_1^{(1)}=\mu_1^{(1)}=1$, and therefore $F=1-\beta^2=1/\gamma^2$.

Away from the source, Eq.~\eqref{eq:app_wave_Ez} is homogeneous. Since the source term for a given azimuthal harmonic is proportional to $\cos(m\vartheta)e^{-jkz}$, linearity allows the field response for that harmonic to be written in the same form,
\begin{equation}
    E_z(r,\vartheta,z;\omega)
    =
    E_{z0}(r)\cos(m\vartheta)e^{-jkz}.
\end{equation}
Substitution into the homogeneous equation gives the radial equation
\begin{equation}
    \frac{1}{r}\frac{d}{dr}
    \left(r\frac{dE_{z0}}{dr}\right)
    -
    \left(\frac{m^2}{r^2}+\nu_i^2\right)E_{z0}=0,
\end{equation}
with
\begin{equation}
    \nu_i
    =
    k\sqrt{1-\beta^2\varepsilon_1^{(i)}\mu_1^{(i)}} .
    \label{eq:app_nu}
\end{equation}
The radial solutions are the modified Bessel functions. Therefore, in region $i$,
\begin{equation}
    E_z^{(i)}
    =
    \left[
    A_i I_m(\nu_i r)+B_i K_m(\nu_i r)
    \right]\cos(m\vartheta)e^{-jkz}.
    \label{eq:app_Ez_general}
\end{equation}
Separation of variables applied to the longitudinal magnetic field, with the angular phase fixed by Maxwell's equations, gives
\begin{equation}
    H_z^{(i)}
    =
    \left[
    C_i I_m(\nu_i r)+D_i K_m(\nu_i r)
    \right]\sin(m\vartheta)e^{-jkz}.
    \label{eq:app_Hz_general}
\end{equation}
The coefficients are fixed by regularity at the origin, decay in the outermost layer, and continuity of the tangential fields at each material interface.

\subsection{Longitudinal impedance}

The longitudinal impedance is obtained from the monopole term $m=0$. Since both sides of the source radius lie in the beam region, the field must be written separately for $r<a$ and $r>a$. We denote these two radial domains by the subscripts $-$ and $+$, respectively. The expressions below are solutions of the homogeneous equation in the open intervals $r<a$ and $a<r<b_1$. The field value at $r=a$ is fixed by continuity of $E_z$, while the radial derivative has a jump determined by the source.

The coefficient $\alpha_{\mathrm{TM}}^{(m)}$ denotes the TM reflection coefficient associated with azimuthal mode $m$. For compactness, define
\begin{equation}
\mathcal F_m(u)
\equiv
K_m(u)-\alpha_{\mathrm{TM}}^{(m)}I_m(u),
\qquad m=0,1 .
\label{eq:app_Fm_def}
\end{equation}
Regularity at the origin eliminates the $K_0$ term for $r<a$, since $K_0(\nu_1 r)$ is singular at $r=0$. Thus
\begin{align}
E_{z,-}^{(1)}(r,z;\omega)
&=
\mathcal A_{-} I_0(\nu_1 r)e^{-jkz},
\qquad r<a,
\label{eq:app_Ez_minus_long}\\
E_{z,+}^{(1)}(r,z;\omega)
&=
\mathcal A_{+}
\mathcal F_0(\nu_1 r)e^{-jkz},
\qquad a<r<b_1 .
\label{eq:app_Ez_plus_long}
\end{align}
Here $\mathcal A_{-}$ and $\mathcal A_{+}$ are constants to be determined.

Defining $x_0\equiv\nu_1 a$, continuity of $E_z$ at $r=a$ gives
\begin{equation}
\mathcal A_{-} I_0(x_0)
=
\mathcal A_{+} \mathcal F_0(x_0).
\label{eq:app_long_continuity}
\end{equation}

The remaining normalization condition determines the amplitudes $\mathcal A_{\pm}$. It is obtained from the discontinuity in the radial derivative at the source. For $m=0$, Eq.~\eqref{eq:app_rho_m} gives
\begin{equation}
    \rho_0(r,z;\omega)
    =
    \frac{Q}{2\pi av}\delta(r-a)e^{-jkz}.
\end{equation}
The radial part of Eq.~\eqref{eq:app_wave_Ez} in the beam region is
\begin{equation}
    \left[
    \frac{1}{r}\frac{\partial}{\partial r}
    \left(r\frac{\partial}{\partial r}\right)
    -
    \nu_1^2
    \right]E_z
    =
    -\frac{jk}{\varepsilon_0}\rho_0 F .
    \label{eq:app_long_radial_inhom}
\end{equation}
Multiplying Eq.~\eqref{eq:app_long_radial_inhom} by $r$ and integrating from $a-\Delta$ to $a+\Delta$, followed by the limit $\Delta\rightarrow0^+$, gives
\begin{equation}
    \left[
    r\frac{\partial E_z}{\partial r}
    \right]_{a^-}^{a^+}
    =
    -\frac{jkQ}{2\pi v\varepsilon_0}F e^{-jkz}.
    \label{eq:app_long_jump}
\end{equation}

Using the fields in Eqs.~\eqref{eq:app_Ez_minus_long} and \eqref{eq:app_Ez_plus_long}, together with the continuity relation Eq.~\eqref{eq:app_long_continuity}, the left-hand side of Eq.~\eqref{eq:app_long_jump} becomes
\begin{equation}
    \mathcal A_{+}
    \frac{x_0}{I_0(x_0)}
    \left[
    I_0(x_0)\mathcal F_0'(x_0)
    -
    \mathcal F_0(x_0)I_0'(x_0)
    \right]e^{-jkz}.
\end{equation}
The terms proportional to $\alpha_{\mathrm{TM}}^{(0)}$ cancel in the bracket. Using the Wronskian identity for the modified Bessel functions at $x_0$,
\begin{equation}
    I_0(x_0)K_0'(x_0)-I_0'(x_0)K_0(x_0)=-\frac{1}{x_0},
\end{equation}
we obtain
\begin{equation}
    \left[
    r\frac{\partial E_z}{\partial r}
    \right]_{a^-}^{a^+}
    =
    -\frac{\mathcal A_{+}}{I_0(x_0)}e^{-jkz}.
\end{equation}
Comparison with Eq.~\eqref{eq:app_long_jump} therefore gives
\begin{equation}
    \mathcal A_{+}
    =
    \frac{jkQ}{2\pi v\varepsilon_0}F I_0(x_0).
\end{equation}
The field for $a<r<b_1$ is then
\begin{equation}
    E_{z,+}^{(1)}(r,z;\omega)
    =
    \frac{jkQ}{2\pi v\varepsilon_0}
    F I_0(x_0)
    \mathcal F_0(\nu_1 r)e^{-jkz}.
    \label{eq:app_Ez_long_final}
\end{equation}

The longitudinal impedance is defined by
\begin{equation}
    Z_\parallel(\omega)
    =
    -\frac{1}{Q}
    \int_0^L E_z(r=a,z;\omega)e^{jkz}\,dz .
\end{equation}
Here $E_z(r=a,z;\omega)$ denotes the continuous value of the field at the source radius, which may be evaluated as the limit of Eq.~\eqref{eq:app_Ez_long_final} as $r\to a^+$. Substituting this expression and using $1/(\varepsilon_0 c^2)=\mu_0$ gives
\begin{equation}
    Z_\parallel^{\mathrm{tot}}(\omega)
    =
    -\frac{jL\omega\mu_0}{2\pi\beta^2}
    I_0^2(x_0)
    \left[
    \frac{K_0(x_0)}{I_0(x_0)}
    -
    \alpha_{\mathrm{TM}}^{(0)}
    \right]F .
    \label{eq:app_Zlong}
\end{equation}
This is Eq.~\eqref{eq:Zlong_final_main} of the main text.

\subsection{Transverse impedance}

The transverse impedance is obtained from the dipole term $m=1$. From Eq.~\eqref{eq:app_rho_m}, the corresponding source term is
\begin{equation}
\rho_1(r,\vartheta,z;\omega)
=
\frac{Q}{\pi av}
\delta(r-a)e^{-jkz}\cos\vartheta .
\end{equation}

The transverse force contains both electric and magnetic contributions, so it is convenient to introduce
\begin{equation}
\mathbf{G}\equiv Z_0\mathbf{H},
\qquad
Z_0=\sqrt{\frac{\mu_0}{\varepsilon_0}} .
\end{equation}

Following the dipole source-field construction used in the multilayer formalism of Ref.~\cite{metralResistivewallImpedanceInfinitely2007}, with additional details given in Refs.~\cite{glucksternAnalyticMethodsCalculating2000,zotterNewResultsImpedance2005}, the longitudinal electric and magnetic fields in the beam region for $a<r<b_1$ are written as
\begin{align}
E_{z,+}^{(1)}(r,\vartheta,z;\omega)
&=
\mathcal B_{+}
\mathcal F_1(\nu_1 r)
\cos\vartheta \,e^{-jkz},
\label{eq:app_Ez_trans}\\
G_{z,+}^{(1)}(r,\vartheta,z;\omega)
&=
\mathcal B_{+}\alpha_{\mathrm{TE}}^{(1)}
I_1(\nu_1 r)
\sin\vartheta \,e^{-jkz}.
\label{eq:app_Gz_trans}
\end{align}
Here $\mathcal B_{+}$ is the dipole source-field amplitude, $\alpha_{\mathrm{TM}}^{(1)}$ is the dipole TM reflection coefficient, and $\alpha_{\mathrm{TE}}^{(1)}$ is an auxiliary TE-like coefficient that appears in the field matching. The normalization follows from the same jump condition as in the longitudinal case, with the factor of two difference coming from the normalization of the $m=1$ azimuthal harmonic. This gives
\begin{equation}
\mathcal B_{+}
=
\frac{jkQ}{\pi v\varepsilon_0}
F I_1(x_0).
\label{eq:app_Bplus_trans}
\end{equation}

The transverse field components can be expressed in terms of the longitudinal ones. In a source-free homogeneous region, the cylindrical components of Faraday's and Ampere--Maxwell's laws give, for a general azimuthal mode $m$,
\begin{align}
E_{r0}
&=
\frac{jk}{\nu_i^2}
\left[
\beta\mu_1^{(i)}\frac{mG_{z0}}{r}
+
\frac{\partial E_{z0}}{\partial r}
\right],
\label{eq:app_Er_from_long}\\
E_{\vartheta0}
&=
-\frac{jk}{\nu_i^2}
\left[
\frac{mE_{z0}}{r}
+
\beta\mu_1^{(i)}
\frac{\partial G_{z0}}{\partial r}
\right],
\label{eq:app_Etheta_from_long}\\
G_{r0}
&=
\frac{jk}{\nu_i^2}
\left[
\beta\varepsilon_1^{(i)}\frac{mE_{z0}}{r}
+
\frac{\partial G_{z0}}{\partial r}
\right],
\label{eq:app_Gr_from_long}\\
G_{\vartheta0}
&=
\frac{jk}{\nu_i^2}
\left[
\frac{mG_{z0}}{r}
+
\beta\varepsilon_1^{(i)}
\frac{\partial E_{z0}}{\partial r}
\right].
\label{eq:app_Gtheta_from_long}
\end{align}
Here $E_{z0}$ and $G_{z0}$ denote the radial amplitudes of the longitudinal fields in region $i$, after the azimuthal and longitudinal dependence has been factored out. In the following, these relations are evaluated in the beam region, $i=1$.

The horizontal transverse impedance is obtained from the Lorentz-force combination
\begin{equation}
E_x-vB_y .
\end{equation}
Since $\mathbf{B}=\mu_0\mu_1^{(1)}\mathbf{H} =\mu_1^{(1)}\mathbf{G}/c$ in the beam region, this can be written as
\begin{equation}
E_x-vB_y
=
E_x-\beta\mu_1^{(1)}G_y .
\end{equation}
For a horizontal displacement, the force can be evaluated in the horizontal plane. Choosing $\vartheta=-\pi/2$, one has $E_x=E_\vartheta$ and $G_y=-G_r$, so that
\begin{equation}
E_x-vB_y
=
E_\vartheta+\beta\mu_1^{(1)}G_r .
\end{equation}
Substituting Eqs.~\eqref{eq:app_Ez_trans} and \eqref{eq:app_Gz_trans} into Eqs.~\eqref{eq:app_Etheta_from_long} and \eqref{eq:app_Gr_from_long}, the terms proportional to $\alpha_{\mathrm{TE}}^{(1)}$ cancel in this force combination. One obtains
\begin{equation}
E_x-vB_y
=
\frac{j\mathcal B_{+}}{kr}
\mathcal F_1(\nu_1 r)e^{-jkz}.
\label{eq:app_lorentz_trans_general}
\end{equation}

At the source radius, this is evaluated as the limit $r\to a^+$,
\begin{equation}
E_x-vB_y
=
\frac{j\mathcal B_{+}}{ka}
\mathcal F_1(x_0)e^{-jkz}.
\label{eq:app_lorentz_trans}
\end{equation}

The transverse impedance is defined as
\begin{equation}
Z_x(\omega)
=
\frac{j}{Qa}
\int_0^L
\left(E_x-vB_y\right)e^{jkz}\,dz .
\end{equation}
Substituting Eq.~\eqref{eq:app_lorentz_trans} and using $Z_0=1/(\varepsilon_0 c)$ gives
\begin{equation}
Z_x^{\mathrm{tot}}(\omega)
=
-\frac{jLZ_0}{\pi\beta}
\frac{I_1^2(x_0)}{a^2}
\left[
\frac{K_1(x_0)}{I_1(x_0)}
-
\alpha_{\mathrm{TM}}^{(1)}
\right]F .
\label{eq:app_Ztrans}
\end{equation}
This is Eq.~\eqref{eq:Ztrans_final_main} of the main text.

\subsection{Field-matching coefficients}

The coefficients $\alpha_{\mathrm{TM}}^{(0)}$ and $\alpha_{\mathrm{TM}}^{(1)}$ encode the response of the surrounding layers and are obtained from the usual cylindrical field-matching conditions \cite{zotterNewResultsImpedance2005,metralResistivewallImpedanceInfinitely2007,MounetThesis}. For the monopole problem, they are obtained by imposing continuity of the tangential fields
\begin{equation}
    E_z^{(i)}=E_z^{(i+1)},
    \qquad
    H_\vartheta^{(i)}=H_\vartheta^{(i+1)}
\end{equation}
at each cylindrical interface. For the dipole problem, the corresponding matching conditions are
\begin{align}
    E_z^{(i)} &= E_z^{(i+1)},
    &
    G_z^{(i)} &= G_z^{(i+1)},
    \nonumber\\
    E_\vartheta^{(i)} &= E_\vartheta^{(i+1)},
    &
    G_\vartheta^{(i)} &= G_\vartheta^{(i+1)} .
\end{align}
Together with regularity at the origin and decay in the outermost region, these conditions determine the reflection coefficients used in the impedance expressions above. For a single external layer, the same boundary conditions give the one-layer expressions quoted in Sec.~II. For multiple surrounding layers, they generate a linear system for the unknown field amplitudes, which is solved to obtain the corresponding values of $\alpha_{\mathrm{TM}}^{(0)}$ and $\alpha_{\mathrm{TM}}^{(1)}$.

\section{\rev{Validation against IW2D for an LHC collimator}}
\label{app:iw2d_validation}

\rev{To further validate the vacuum limit of the present formulation, we compare the transverse wall impedance with independent IW2D calculations for an LHC graphite collimator, both without and with a thin copper coating. Following Ref.~\cite{metralTransverseResistivewallImpedance2005}, we consider a cylindrical structure of length $L=1\,\mathrm{m}$ and inner radius $b_1=2\,\mathrm{mm}$, with a graphite resistivity of $\rho_{\mathrm{g}}=10\,\mu\Omega\,\mathrm{m}$. In the coated configuration, a copper layer of thickness $5\,\mu\mathrm{m}$ and resistivity $\rho_{\mathrm{Cu}}=17\,\mathrm{n}\Omega\,\mathrm{m}$ is introduced between the vacuum beam region and the graphite. The Lorentz factor is $\gamma=7462.69$. Both materials are modeled with frequency-independent conductivity and relative permittivity and permeability equal to unity. In the present formulation, the source radius is set to $a=0.1\,\mu\mathrm{m}$ to approximate the infinitesimal-offset limit used by IW2D.
}

\begin{figure}[!htbp]
    \centering
    \includegraphics[width=\linewidth]{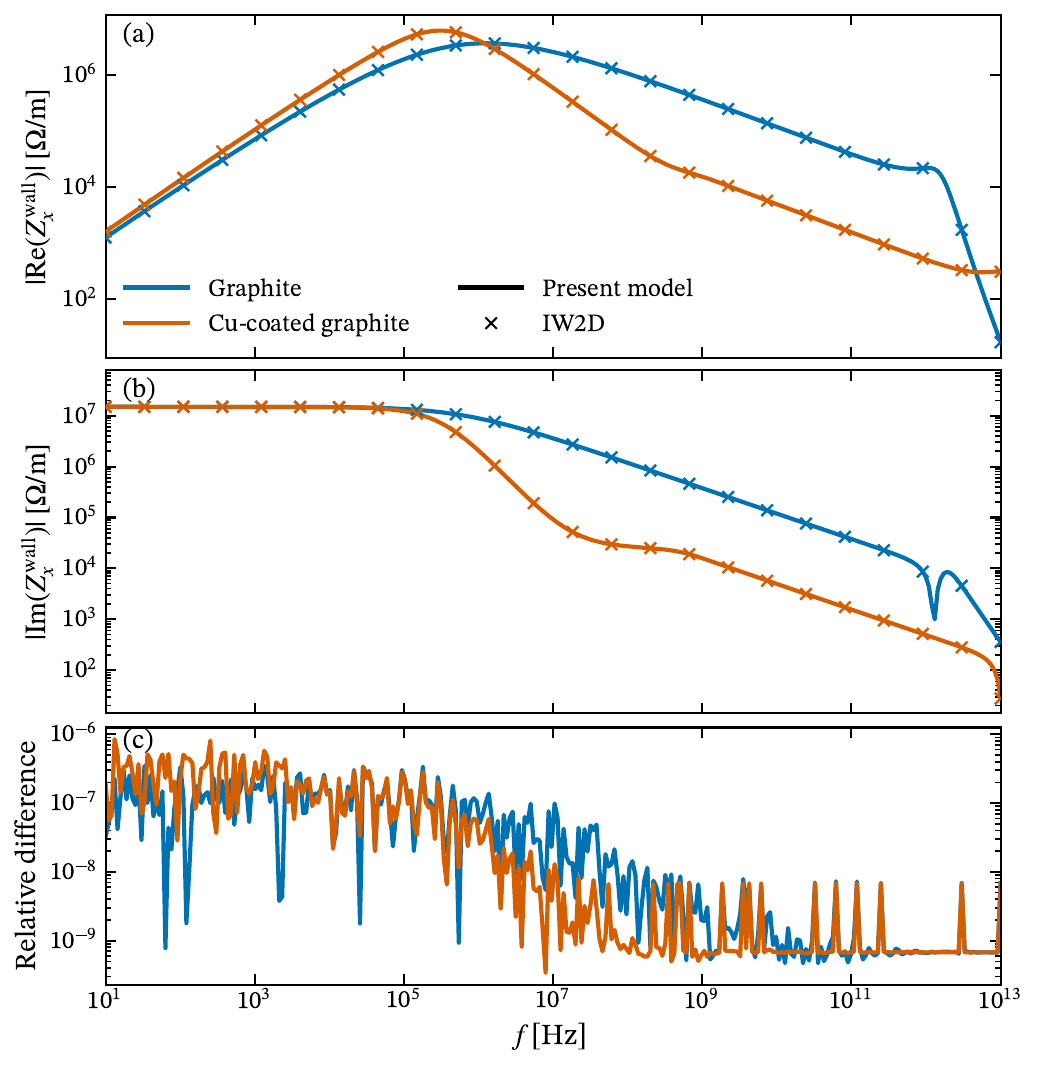}
    \caption{\rev{Comparison of the transverse wall impedance calculated using the present formulation (solid lines) and IW2D (crosses) for an LHC graphite collimator, with and without a copper coating. Panels (a) and (b) show the magnitudes of the real and imaginary parts, respectively, while panel (c) shows the relative difference of the complex impedance.}}
    \label{fig:iw2d_collimator}
\end{figure}

\rev{Figure~\ref{fig:iw2d_collimator} shows excellent agreement between the present formulation and IW2D for both configurations over the frequency range considered.}

\FloatBarrier

\bibliography{apssamp}

\end{document}